\documentclass[twocolumn,iop,apj]{aastex63}
\usepackage{amsfonts,amsmath,graphicx,natbib,url,hyperref,nicefrac}
\usepackage{amssymb, verbatim, subfigure, paralist, soul, comment}
\usepackage{graphicx}
\usepackage{wrapfig}
\usepackage{verbatim}
\usepackage{lineno}
\usepackage{tikz}
\usepackage{color}

\usepackage{overpic}
\usepackage{multirow}
\usepackage{float}

\usepackage[T1]{fontenc}
\usepackage{aecompl}
\usepackage{calligra}
\DeclareMathAlphabet{\mathcalligra}{T1}{calligra}{m}{n}
\DeclareFontShape{T1}{calligra}{m}{n}{<->s*[2.2]callig15}{}

\usepackage{ulem}
\usepackage{cancel}
\usepackage{hyperref}
\hypersetup{
    pdfnewwindow=true,      % links in new window
    colorlinks=true,       % false: boxed links; true: colored links
    linkcolor=blue,          % color of internal links
    citecolor=blue,        % color of links to bibliography
    filecolor=blue,      % color of file links
    urlcolor=blue           % color of external links
}

\def\apjl{ApJL}               % Astrophysical Journal, Letters
\def\apjs{ApJS}               % Astrophysical Journal, Supplement
\def\aap{A\&A}                % Astronomy and Astrophysics
\DeclareMathAlphabet{\mathcalligra}{T1}{calligra}{m}{n}
\DeclareFontShape{T1}{calligra}{m}{n}{<->s*[2.2]callig15}{}

\def\eg{{\em e.g.}}

\hypersetup{colorlinks=true, urlcolor=blue, citecolor=blue}

\usepackage{comment}

\shorttitle{} 
\shortauthors{O'Neill et al.}

\begin{document}

\title[]{Gravitational Wave Decoupling in Retrograde Circumbinary Disks}

\author[0000-0002-1382-3802]{David O'Neill}
\affiliation{Niels Bohr International Academy, Niels Bohr Institute, Blegdamsvej 17, DK-2100 Copenhagen Ø, Denmark}
\email{david.oneill@nbi.ku.dk}

\author[0000-0002-3820-2404]{Christopher Tiede}
\affiliation{Niels Bohr International Academy, Niels Bohr Institute, Blegdamsvej 17, DK-2100 Copenhagen Ø, Denmark}
% \email{christopher.tiede@nbi.ku.dk}

\author[0000-0002-1271-6247]{Daniel J. D'Orazio}
\affiliation{Space Telescope Science Institute, 3700 San Martin Drive, Baltimore , MD 21}
\affiliation{Niels Bohr International Academy, Niels Bohr Institute, Blegdamsvej 17, DK-2100 Copenhagen Ø, Denmark}
%\email{dorazio@stsci.edu}

\author[0000-0003-3633-5403]{Zolt\'an Haiman}
\affiliation{Department of Astronomy, Columbia University, 550 West 120th Street, New York, NY, 10027, USA}
\affiliation{Department of Physics, Columbia University, 550 West 120th Street, New York, NY, 10027, USA}
\affiliation{Institute of Science and Technology Austria (ISTA), Am Campus 1, Klosterneuburg 3400, Austria}

\author[0000-0002-0106-9013]{Andrew MacFadyen}
\affiliation{Center for Cosmology and Particle Physics, Physics Department, New York University, New York, NY 10003, USA}

\begin{abstract}
We present a study of the late-time interaction between supermassive black hole binaries and retrograde circumbinary disks during the period of gravitational wave-driven inspiral. While mergers in prograde disks have received extensive study, retrograde disks offer distinct dynamics that could promote mergers and produce unique observational signatures. Through numerical simulations, we explore the process of binary-disk decoupling, where the binary's orbital decay rate is faster than the disk's viscous response rate. We find the point of decoupling to be comparable in prograde and retrograde disks, suggesting that any associated electromagnetic (EM) signatures will be produced at comparable times preceding merger. However, we find smaller central cavities for retrograde disks, likely leading to higher-frequency EM emissions and shorter post-merger rebrightening timescales compared to their prograde counterparts. Additionally, we identify quasi-periodic flaring due to instabilities unique to low-viscosity retrograde disks, which may produce distinctive EM signatures.\\
\end{abstract}

\section{Introduction}
The gravitational waves (GWs) produced by compact supermassive black hole binaries (SMBHBs) propagate through the universe, encoded with information of the binary's final moments before merger. The advent of next-generation space-based GW detectors (\eg~the Laser Interferometer Space Antenna (LISA) \citealt{2023LRR....26....2A}, the Taiji program \citealt{2020IJMPA..3550075R} and TianQin \citealt{2021NatAs...5..881G}) will enable such observations. Accompanied by theoretical modelling, these detections may reveal the nature of their astrophysical environments \citep{Amaro_Seoane_2023}, fundamental physics in the strong gravity regime \citep{afroz2024modelindependentprecisiontestgeneral, EMRI_Tests_of_GR_Speri} and the cosmological history of massive mergers \citep{Auclair2023CosmologyWT}.\\

GWs alone are inefficient at producing compact SMBHBs. Instead, the external environment is required to reduce the binary separation from the galactic scale $\sim$ kpc, to the sub-pc gravitational wave inspiral scale \citep[eg.][]{1980Natur.287..307B, Khan_2012, Kelley2017}. Following the merger of two massive galaxies, the two remnant cores will sink to the center of the newly formed galaxy via dynamical friction \citep{Chandrasekhar_DF, GadgetSimulations, DiMatteo2023}, where they will encounter each other and become bound. Stars with orbits intersecting this binary (so called `loss-cone orbits') can experience a gravitational slingshot, thereby extracting orbital energy from the binary, further reducing its separation \citep{2006ApJ...642L..21B, 2009ApJ...695..455B}. This process will slowly tighten the SMBHB until around $\lesssim 0.1\mathrm{pc}$ \citep{Armitage2002AccretionDT} from which point many of these binaries are expected to interact with ambient gas in the galactic nucleus and form a circumbinary accretion disk \citep{1991ApJ...370L..65B,Artymowicz_1996,2008ApJ...672...83M,Tanaka2013ElectromagneticSO, Mayer_2013}. This disk will exert a force on the binary, changing its orbital elements \citep{Armitage2002AccretionDT, GouldRix2000, Moody_2019,2021ApJ...914L..21D, 2021ApJ...909L..13Z} until gravitational waves become sufficient to drive the binary to merge. These accretion disks are of particular interest because they can source bright multi-wavelength EM radiation \citep{2012ApJ...761...90G, 2012MNRAS.427.2680K,Westernacher_Schneider_2022,2023arXiv231016896D} with the possibility of remaining bright all the way until merger \citep{10.1093/mnrasl/slu184,FarrisDuffel2015, 10.1093/mnras/stad3095,Franchini2024}. For this reason, an observable EM counterpart may accompany a SMBHB merger \citep{Schnittman_2008,Bowen_2018,Tang_2018,Paschalidis_2021,Bogdanovi__2022, Wang_2022}.\\

An EM signal coincident with a gravitational wave measurement would greatly enhance the significance of the detection \citep{Holz_2005,Tamanini_2016,Mangiagli_2022}. Complementary to gravitational waves, EM signals would provide much better sky localisation, possibly enabling a unique host galaxy to be identified with each event -- allowing independent measurements of the merger mass and redshift via their spectra and broad-line regions \citep[\eg,][]{casura2024exploringmassmeasurementssupermassive}. Time-domain surveys of the host galaxies may capture the expected inspiral signatures, such as an abrupt disappearance of X-rays and dimming in UV \citep[as seen in][]{10.1093/mnras/stad3095, 2024arXiv240510281C, Franchini2024} followed by their gradual re-brightening. Moreover, EM signatures alone could enable indirect detections of SMBHB mergers when gravitational wave detectors are offline or when systems are outside of the detectors' sensitivity range. To quantify the signatures produced, we study circumbinary disks hosting a gravitational wave inspiral.\\

In a steady-state, the angular momentum of an accretion disk can be either aligned (prograde) or misaligned (retrograde) with the angular momentum vector of the binary. Prograde disks have received extensive attention \citep[eg.][along with all previously referenced circumbinary disk studies]{2008ApJ...672...83M,10.1111/j.1365-2966.2008.14147.x, 10.1093/mnras/stt1787} whereas retrograde ones much less so.  Depending on the nature of the accretion episodes, the retrograde configuration is not only stable, but almost as likely %but has been argued to be comparably likely 
to form as the prograde one \citep{10.1111/j.1745-3933.2006.00249.x,10.1111/j.1365-2966.2012.21072.x,2013ApJ...774...43M,Bankert_2015}. We argue that retrograde disks may be a promising environmental source of SMBHB mergers for a number of reasons:
\begin{itemize}
\item Irrespective of orbital parameters, retrograde disks always facilitate orbital inspiral. In contrast, prograde disks have been observed to produce orbital outspirals under specific disk conditions and binary orbital parameters \citep{Muñoz_2019,2021ApJ...914L..21D, Siwek2023}.
\item The torque experienced by a binary in a retrograde disk is stronger than in a prograde counterpart \citep[][and references therein]{Tiede_2023}, facilitating a faster orbital decay.
\item Unlike prograde disks, eccentricity is always driven in retrograde disks \citep{Nixon2011, Tiede_2023}. Larger eccentricities will lead to an earlier onset of the gravitational wave dominated regime, thereby promoting SMBHB mergers.
\item Large initial eccentricities may not completely circularise before entering detector sensitivity, leaving direct observable consequences \citep{2024MNRAS.528.4176G, DeLaurentiis2024}.
\end{itemize}

In this paper, we focus on the process of \emph{binary-disk decoupling}-- defined as the point at which the binary's orbital decay due to gravitational radiation outpaces the disk's ability to viscously react. After decoupling, the binary quickly contracts toward merger, leaving the circumbinary disk behind. Depending on the viscosity of the disk, decoupling can occur at different binary semi-major axes \citep{Armitage2002AccretionDT} and, therefore, at different times preceding the merger. For a range of different binary masses ($10^4\to10^7 M_\odot$) and typical values for viscosity, \cite{Dittmann_2023} found that high-viscosity prograde disks likely decouple within the LISA band (see their figure 5), enabling coincident EM and gravitational wave detections of the event. However, despite retrograde disks potentially accounting for a significant portion of SMBHB mergers, decoupling in retrograde disks has remained unexplored. To address this, we perform numerical simulations of the late-time interaction between an equal-mass binary inspiralling within a retrograde circumbinary disk. We focus on (a) the process of decoupling in retrograde disks and (b) the disk dynamics throughout the inspiral, thereby helping to inform the associated EM counterpart.

\section{Methods}
\subsection{Orbital Dynamics}
\label{sec:OrbitalDynamics}
We begin by describing the orbital dynamics of a gravitational wave inspiral. While retrograde disks promote eccentricity growth, we assume that the binary is initially circular ($e=0$) as a first step to simplify the key dynamics involved in the system. In addition, we assume the binary to be of equal-mass components (with total mass $M$) and coplanar with the disk ($\iota = 0$). Under these assumptions, the orbital phase $\phi$ and separation $a$ uniquely determine the binary dynamics. Correspondingly, we adopt the initial semi-major axis $a_0$ as unit length, along with the initial angular frequency $\Omega_0 = \sqrt{GM/a_0^3}$ as inverse unit time. Following \cite{1964PhRv..136.1224P}, we assume that the orbital energy changes over a timescale much longer than the orbital period, permitting an approximation to the quadrupolar formula \citep{https://doi.org/10.1002/andp.19163540702} as a series of closed Keplerian orbits. In doing so, the semi-major axis $a(t)$ and orbital phase $\phi(t)$  become time-dependent quantities,
\begin{align}
	\frac{a(t)}{a_0} &= \left(1-\frac{64}{5}\frac{G^3M^3}{a_0^4c^5}t\right)^\frac{1}{4}\label{SemiMajorAxisInspiral}\\
	\phi(t) &= \int_0 ^ t dt' \sqrt{\frac{GM}{a(t')^3}}.
 \label{PhaseChange}
\end{align}
Here $c$ is the speed of light, which, when expressed in the unit system above $c=\tilde{c} ~ a_0\Omega_0$,
\begin{equation}
	\tilde{c} = \sqrt{\frac{c^2 a_0}{GM}} = \sqrt{\frac{a_0}{r_\mathrm{G}}},
 \label{SpeedOfLight}
\end{equation}
where $r_G$ is the gravitational radius. The numerical value of $\tilde{c}$ determines the timescale for orbital inspiral. We adopt a fiducial value of $a_0 = 100 r_\mathrm{G}$ corresponding to a merger timescale of $t_\mathrm{m}\approx 1244\times 2\pi/\Omega_0$ (approximately $1990$ binary orbits). Finally, we neglect relativistic corrections to fluid motion near the black holes and treat the binary components as Plummer potentials,
\begin{equation}
    \Phi_b(\mathbf{x}) = -\frac{GM}{2\sqrt{|\mathbf{x}-\mathbf{x}_1|^2+r_\mathrm{soft}^2}} - \frac{GM}{2\sqrt{|\mathbf{x}-\mathbf{x}_2|^2+r_\mathrm{soft}^2}},
    \label{BinaryPotential}
\end{equation}
where $\mathbf{x}_1, \mathbf{x}_2$ are the current positions of the binary components and $r_\mathrm{soft}$ is the gravitational softening length.

\subsection{Disk Dynamics}
We model the disk in 2D as a viscous fluid in the presence of the time-changing gravitational potential of the binary ($\Phi_b$). Hence we solve the vertically integrated mass conservation and Navier-Stokes equations for disk surface density $\Sigma$ and mid-plane fluid velocity $v^i$,
\begin{align}
	\partial_t \Sigma +\partial_i \left(\Sigma v^i\right) &= S_\Sigma \label{MassContinuity}\\
	\partial_t(\Sigma v^j) + \partial_i \left(\Sigma v^iv^j + P\delta^{ij} - \tau^{ij} \right) &= S^j -\partial^j\Phi_\mathrm{b},
  \label{NavierStokes}
\end{align}
where $P$ is the $2D$ pressure, $\delta^{ij}$ is the Kronecker delta and $S_\Sigma, S^j$ are the mass and momentum sinks. The viscous stress tensor
\begin{equation}
	\tau^{ij} = \nu\Sigma\left(\partial^iv^j + \partial^jv^i-\frac{2}{3}\delta^{ij}\partial_kv^k\right),
 \label{ViscousStress}
\end{equation}
describes the transport of momentum $\Sigma v^i$ across surfaces of constant $j$, with $\nu$ the kinematic viscosity of the fluid. 
We assume a locally isothermal equation of state by prescribing the sound speed $c_\mathrm{s}$ as
\begin{equation}
    c_\mathrm{s}^2 = -\frac{\Phi_b}{\mathcal{M}^2},
    \label{SoundSpeed}
\end{equation}
for which the vertically integrated pressure is given by $P = \Sigma c_\mathrm{s}^2$. Here, $\mathcal{M}$ is the constant Mach number in the disk. Larger values of $\mathcal{M}$ correspond to colder, thinner disks with $\mathcal{M} = r/H$ in vertical hydro-static equilibrium. We adopt a fiducial value $\mathcal{M}=10$ for all simulations.\\

\subsection{Decoupling}
We parameterise viscous angular momentum transport through the disk with the coefficient of kinematic viscosity, $\nu$. This angular momentum flux leads to an inward flow of mass with radial velocity
\begin{equation}
	v_\mathrm{r} = -\frac{3\nu}{2r}\left(1+2\frac{r}{\nu r}\frac{d}{dr}{\nu\Sigma}\right).
 \label{ViscousVelocity}
\end{equation}
Eq.~\ref{ViscousVelocity} sets a `viscous rate' determining the timescale required for gas to re-arrange its angular momentum and re-orient itself across the disk. An analytical approximation for binary-disk decoupling can be obtained by equating the rate of change of binary semi-major axis (the derivative of Eq.~\ref{SemiMajorAxisInspiral}) with the radial velocity due to viscosity in Eq.~\ref{ViscousVelocity}. In this estimate, the moment at which they are equal, is the final time that the disk is viscously coupled to the inspiral of the binary. We make two assumptions regarding the viscous rate from Eq.~(\ref{ViscousVelocity}): (1) that $\nu\Sigma = \dot{M}/3\pi$ is a constant, while, (2) the circumbinary disk has a depleted central region (otherwise known as a cavity) between the barycenter and some typical radius $r = \xi a_b$. It is at this radius (not necessarily the binary semi-major axis) that we measure the radial velocity of the gas. Therefore, the semi-major axis of the binary at decoupling $a_\mathrm{dc}$ is,
\begin{equation}
    a_\mathrm{dc} = \sqrt{\frac{32\xi}{15\nu}\frac{G^3M^3}{c^5}}.
    \label{Decoupling}
\end{equation}
This velocity-based argument for decoupling agrees with numerical simulations of inspirals in prograde disks \citep[see Figure 2 of][]{Dittmann_2023}.

\begin{figure*}[t]
    \centering
    \includegraphics{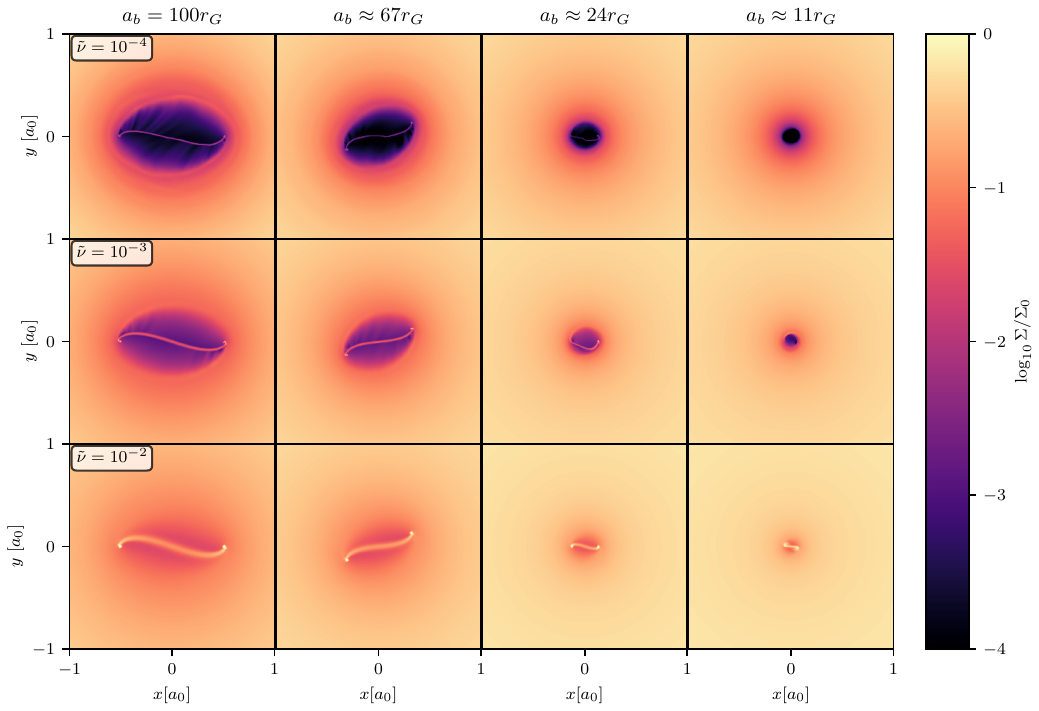}
    \caption{Logarithmic snapshots of the disk surface density $\Sigma/\Sigma_0$ across a range of binary semi-major axes (columns) and viscosities (rows). The left-most panels illustrate the disk structure following the viscous relaxation period of $1000$ binary orbits -- preceding the onset of a gravitational wave inspiral. As the inspiral progresses the semi-major axis of the binary decreases (right) and the disk reacts, primarily through a shrinking central cavity. Decoupling happens earlier for less viscous disks (top) while the intrabinary bridge is more stable for larger viscosities (bottom).}
    \label{fig:DiskMorphology}
\end{figure*}

\subsection{Numerical Methods}
All simulations were performed using the publicly-available, GPU-accelerated code, $\texttt{Sailfish} $ (for further details see \citealt{2024ascl.soft08004Z}, and \citealt{2024ApJ...970..156D} for a code comparison). $\texttt{Sailfish}$ is a second-order, Newtonian, grid-based hydrodynamics code designed to solve Eqs.~(\ref{MassContinuity}) and (\ref{NavierStokes}) numerically on a fixed Cartesian grid. Below, we discuss our main numerical methods.\\

The computational domain is a two-dimensional Cartesian grid, of size $20a_0 \times 20a_0$ with resolution $n = 3000 \times 3000$. Centered on this domain is a circular outer-boundary `buffer region' of radius $9a_0$ -- outside of which, the solution is damped back to that of a steady-state, axisymmetric accretion disk. The softened Plummer potentials Eq.~(\ref{BinaryPotential}) represent inner boundaries wherein mass and momentum sinks $S_\Sigma, S^j$ emulate the accretion of material onto the binary. We use an acceleration-free sink prescription \citep[][]{DittmannRyan2021} in modelling a physical horizon-like boundary which removes mass and momentum at a rate of $\gamma_\mathrm{sink} = 50\Omega_0$ (for further discussion see Section~\ref{sec:Minidisks}), with radius equal to the Innermost Stable Circular Orbit (ISCO) $r_\mathrm{sink} = 0.03 a_0$ (and softening length $r_\mathrm{soft}=0.03 a_0$, unless specified otherwise). The cell spacing is $\delta \approx 0.0067 a_0$, meaning that there are $\sim 63$ cells covering the area of the sink.\\

We initialise each simulation with a uniform density profile $\Sigma/\Sigma_0 = 1$ and allow the disk to viscously relax for $1000$ binary orbital periods prior to the onset of a gravitational wave inspiral. During this relaxation period, we implement a gentle-sink for the first $800$ binary orbits with a rate $\gamma_\mathrm{sink} = 1\Omega_0$. We find this gentle-sink period to be necessary for the numerical stability of the solution at the beginning of the simulation. We perform both prograde and retrograde simulations, which are identical other than the initial azimuthal flow of the gas over a range of different kinematic viscosities, $\nu = \tilde{\nu} ~ a_0^2 \Omega_0$ where $\tilde{\nu} \in \{10^{-4}, 3\times10^{-4}, 10^{-3}, 3\times10^{-3}, 10^{-2}\}$. \footnote{These constant kinematic viscosities correspond to $\alpha$-viscosity models \citep{1973A&A....24..337S} of $\alpha \in [10^{-2}, 3\times10^{-2}, 10^{-1}, 3\times10^{-1}, 1]$ when evaluated at the binary semi-major axis.}

\section{Results}

\begin{figure*}[t]
    \centering
    \includegraphics[width=\textwidth]{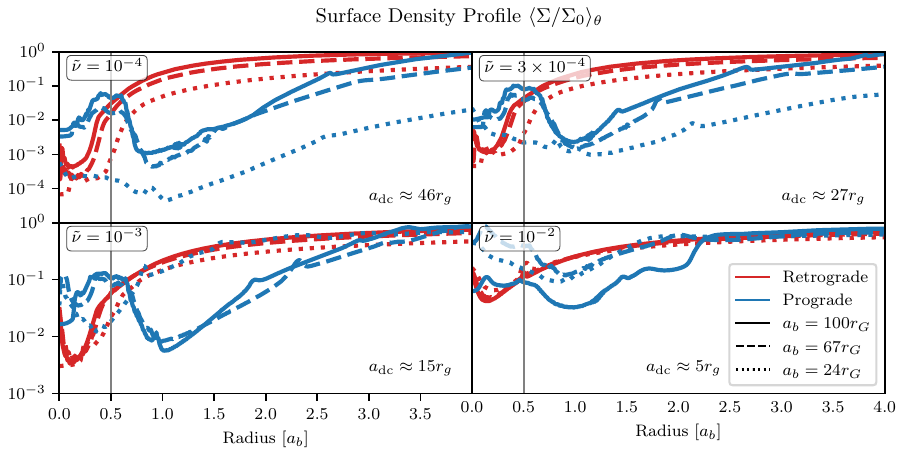}
    \caption{
    Azimuthally averaged surface densities for retrograde disks (red) and prograde disks (blue) at three different stages during the inspiral (solid, dashed and dotted). The x-axis units have been scaled to the current orbital separation of the binary (grey vertical lines). On each panel we print the nominal decoupling semi-major axis $a_\mathrm{dc}$ (see Eq.~\ref{Decoupling} with $\xi = 1$).
    }
    \label{fig:AverageSurfaceDensity}
\end{figure*}

\subsection{Disk Morphology}
\label{sec:DiskMorphology}
During the inspiral, the disk evolves in response to the binary. We illustrate this evolution through 2D surface density profiles in Fig.~\ref{fig:DiskMorphology} and the azimuthally averaged density profiles in Fig.~\ref{fig:AverageSurfaceDensity} for a selection of different times and viscosities.\\

Perhaps the most prominent feature of a circumbinary disk is the central cavity, a low density region of the disk, excavated by the binary. In the left panels of Fig.~\ref{fig:DiskMorphology}, we approximate the cavity (see Appendix~\ref{Appendix}) as an ellipse with semi-major axis $a_\mathrm{c} = a_0$ and eccentricity $e_\mathrm{c} \approx 0.65$ centered and orientated with the binary. The cavity semi-major axis $a_\mathrm{c} = a_0$ is significantly smaller than typical prograde values between $2a_0$ and $5a_0$ \citep{Hirsh2020}, suggesting that a retrograde circumbinary disk can remain stable much closer to the binary than a prograde disk can (for further discussion, see Section \ref{sec:RetrogradeStability} on the stability of retrograde orbits). The cavity orientation locks with the binary before the inspiral \citep{Tiede_2023} and during the early stages of the inspiral (left panels of Fig.~\ref{fig:DiskMorphology}), while at later times (as the binary approaches merger), the cavity drops in eccentricity, becoming more axisymmetric meaning that information on the cavity’s orientation is lost (right panels of Fig.~\ref{fig:DiskMorphology}). In contrast, lopsided, eccentric cavities persist during inspiral for prograde configurations as seen in Figure 1 of both \cite{Dittmann_2023} and \cite{10.1093/mnras/stad3095}.\\ 

In Fig.~\ref{fig:AverageSurfaceDensity} we illustrate the azimuthally averaged surface density profiles $\langle\Sigma/\Sigma_0\rangle_\phi$ across a range of viscosities and times. We draw comparisons to reference prograde disks (blue) to better understand the qualitative differences in disk structure throughout the inspiral. Retrograde disks maintain a larger surface density closer to the binary, virtue of a smaller cavity -- with the notable exception of the prograde minidisks (the blue bumps near $r = 0.5 a_b$). As the viscosity increases (lower right panel), the disk becomes increasingly capable of retaining its initial surface density profile throughout the inspiral. Retrograde disks also exhibit less variability in surface density with viscosity and time than their prograde counterparts.\\

\begin{figure}[t] 
    \centering
    \includegraphics[width=0.92\linewidth]{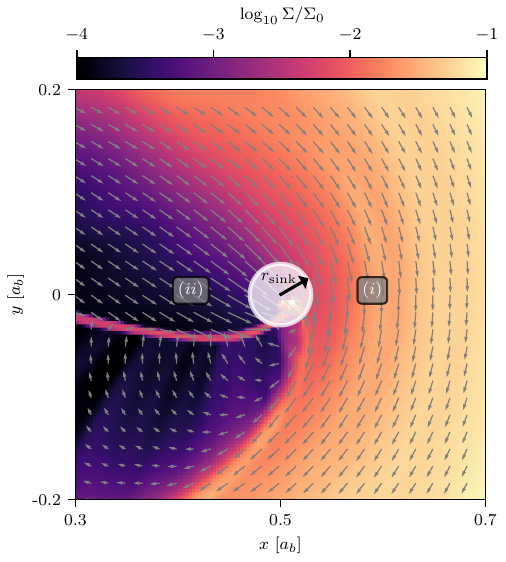}
    \caption{Zoom in on one of the binary components. The arrows represent the direction of fluid velocity in the frame corotating with the binary and the colourmap is the logarithmic surface density. Regions $(i)$ and $(ii)$ experience significant deflections by the binary and collide, forming a shock behind each binary component. }
    \label{fig:BridgeFormation}
\end{figure}

An intrabinary bridge is present in a selection of panels from Fig.~\ref{fig:DiskMorphology}. As the binary rotates along its orbit, it approaches circumbinary disk material which is \emph{exterior} to the binary orbit (region ($i$) of Fig.~\ref{fig:BridgeFormation}) and cavity material, which is interior to the binary orbit (region ($ii$) of Fig.~\ref{fig:BridgeFormation}). Particularly close encounters between the gas and the binary result in the deflection of material, although in directions opposite for regions ($i$) and ($ii$). As a result, two opposing streams of material collide behind each binary component and a shock is formed. The locations at which these shocks occur constitutes the intrabinary bridge with a characteristic `S' shape. Material on this bridge has low angular momentum and can be transported almost radially, before being accreted. In Section \ref{sec:BridgeInstabilities} we discuss the stability of the intrabinary bridge for low-viscosity retrograde disks.

\subsection{Minidisks}
\label{sec:Minidisks}
The lack of circum-single minidisks during retrograde simulations is evident in Figs.~\ref{fig:DiskMorphology}-\ref{fig:AverageSurfaceDensity}. Previous studies have debated their existence, with \cite{Tiede_2023} and \cite{10.1111/j.1365-2966.2012.21072.x} finding persistent minidisks whereas \cite{10.1093/mnras/stu194} did not. During our retrograde simulations, we observe minidisks only during the `gentle sink' relaxation phase.\\

\begin{figure}[H] %[H]
    \centering
    \includegraphics[width=\linewidth]{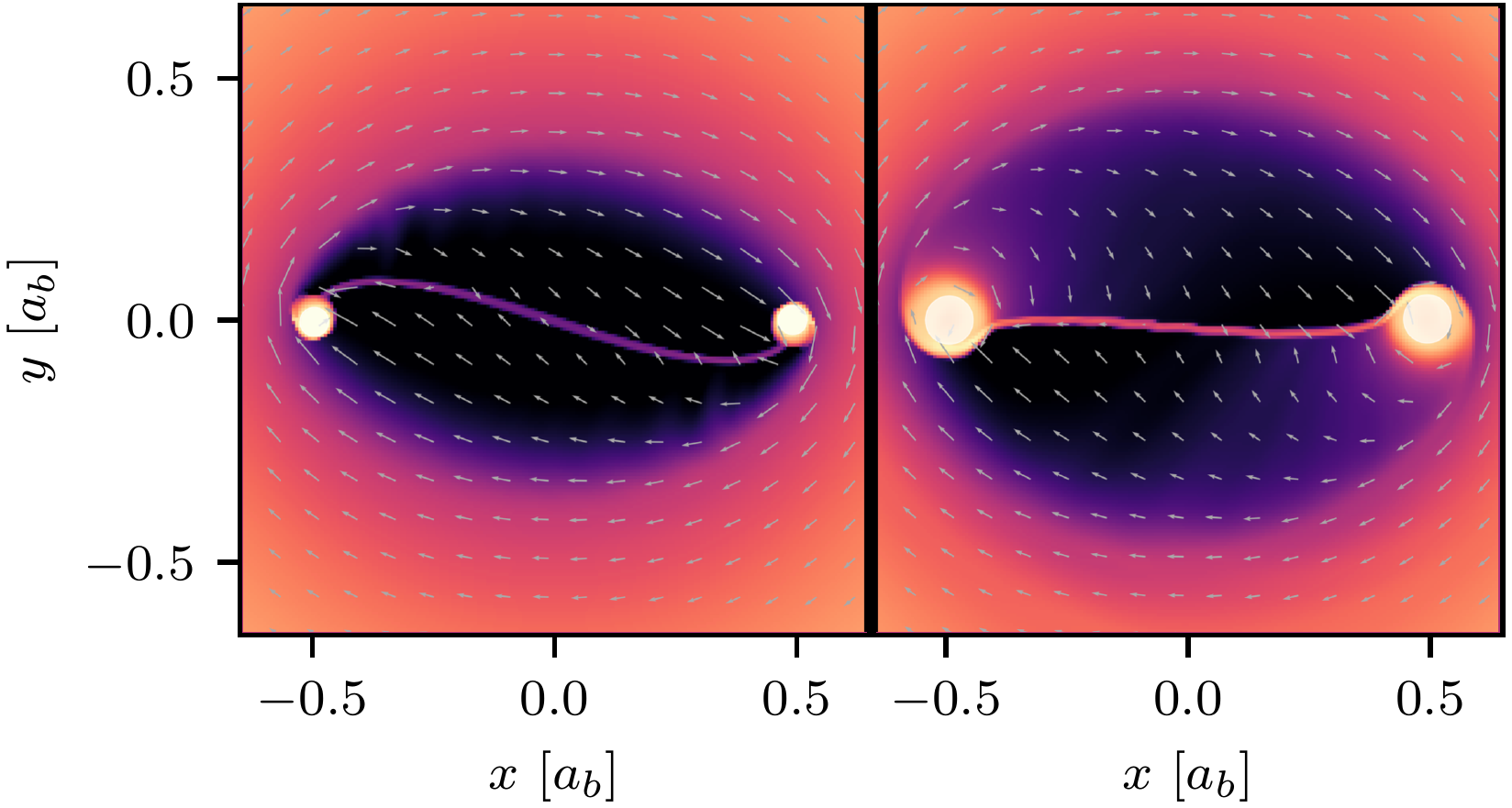}
    \caption{Two stable minidisk configurations for $\gamma_\mathrm{sink} = 1\Omega_0$, corresponding to different values of the softening radius, $r_\mathrm{soft} = 0.03a_0$ (left) and $r_\mathrm{soft} = 0.05a_0$ (right) in otherwise identical simulations. These density snapshots are taken at $t = 200 [2\pi\Omega_0^{-1}]$ and the colorbar is the same as Fig.~\ref{fig:DiskMorphology}.
    }
    \label{fig:Configurations}
\end{figure}

For a low sink rate ($\gamma_\mathrm{sink} = 1\Omega_0$) we observe two different stable solutions determined by the softening radius $r_\mathrm{soft}$, as illustrated in Fig.~\ref{fig:Configurations}. On the one hand, a large softening radius ($r_\mathrm{soft} = 0.05 a_0$ right panel) results in a shallow gravitational potential, allowing bound material to spread over a wider area, resulting in broader minidisks\footnote{Where the softening radius is assumed to be less than the radius of the Roche-bound region $r_\mathrm{R} \approx 0.15 a_b$. For further discussion, see Section~\ref{sec:RetrogradeStability}.}. As these wide minidisks rotate through the cavity, substantial ram-pressure generates shocks which cause the intrabinary bridge to attach to the front of the minidisk, rather than to the rear. On the other hand, a small softening radius results in smaller minidisks, which are attached to the rear by the intrabinary bridge (left). We verified that larger sink radii ($r_\mathrm{sink} = 0.05a_0$) have no effect on the size of the minidisks or the structure of the intrabinary bridge.\\

The bridge is the funnel through which material is channeled onto the minidisk. Front-attaching bridges (right panel of \ref{fig:Configurations}) source \emph{retrograde} minidisks, whereas rear-attaching bridges (see left panel of Fig.~\ref{fig:BridgeFormation}) source \emph{prograde} minidisks. Therefore, for a low sink rate ($\gamma_\mathrm{sink} = 1\Omega_0$) the size of the minidisks, their sense of rotation and the intrabinary bridge connecting them can be in different, stable configurations depending on the softening radius.\\

We note that prograde circumbinary disks host larger minidisks than retrograde circumbinary disks -- with typical radii of $\mathcal{R}_\mathrm{P} \approx 0.3a_0$ for the former and $\mathcal{R}_\mathrm{R} \approx 0.1a_0$ for the latter (right panel Fig.~\ref{fig:Configurations}). The origin of this difference can be attributed to different effective Roche-bound radii and will be discussed in Section~4.1 below.\\

\begin{figure*}
    \centering
    \vspace{-2mm}
    \includegraphics[width=\textwidth]{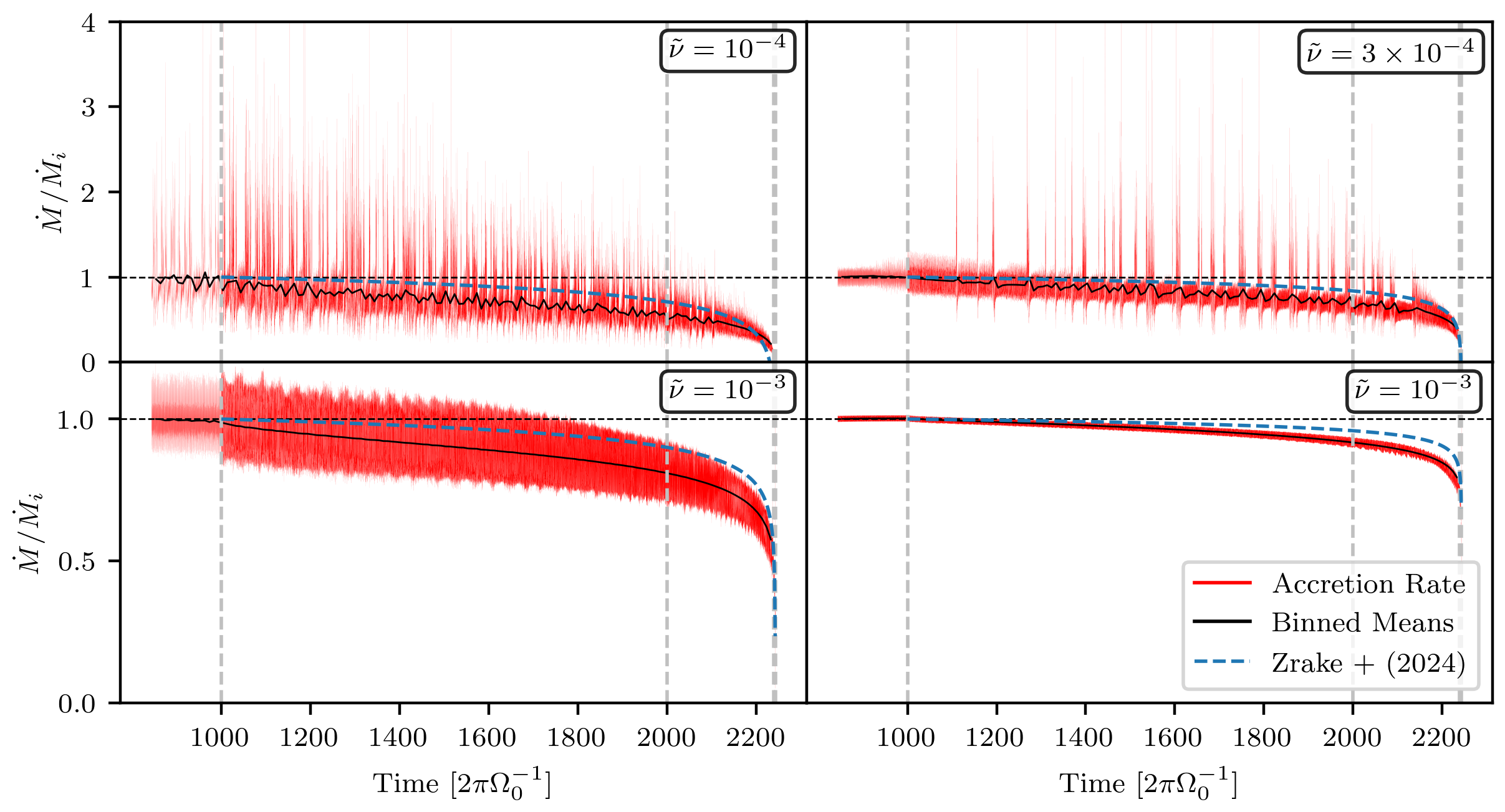}
    \caption{Total accretion rates $\dot{M} = \dot{M}_1 + \dot{M}_2$ of retrograde binaries for the viscosities noted in the top right of each panel. The inspiral is initiated at $t = 1000$ orbits, at which time the timeseries cadence is also increased. The dashed-grey lines corresponding to the times at which the density snapshots are taken in Fig.~\ref{fig:DiskMorphology} (ie. $a_b = [100r_G, 67r_G, 24r_G, 11r_G]$) in all panels. The accretion rate is normalised to unity in each panel, with $\dot{M}_\mathrm{i}> \dot{M}_0$ (this overshoot is discussed in \cite{2024arXiv240510281C} and attributed to an artifact of the outer boundary). The dashed blue lines represent the ``weakening-torque effect'' from \cite{zrake2024changinglookinspiralstrendsswitches}.}
    \label{fig:AccretionRates}
\end{figure*}

At the scales resolvable for an inspiralling binary, the sink represents the ISCO, within which all material must be accreted to emulate a physical boundary. Therefore, the sink rate must be sufficiently high to ensure that once material crosses the inner boundary it is swiftly removed from the domain to prevent artificial pile-ups. Irrespective of the softening radius of the disks, when a faster sink is implemented ($\gamma_\mathrm{sink} = 50 \Omega_0$), the resulting solution always features a rear-attaching bridge \emph{without circumsingle minidisks} -- as depicted in Fig.~\ref{fig:BridgeFormation}. For completeness, we verified that higher sink rates have no effect on the simulation results of prograde disks.

\subsection{Accretion and Torque Timeseries}

The total accretion rates ($\dot{M}$) along with the torque exerted on the disk\footnote{$\mathcal{T} = - a_b \times \mathbf{F}$ where $\mathbf{F}$ is the force exerted on the binary which is comprised of both a gravitational $\mathbf{F_g}$ and accretion $\mathbf{F}_a$ component.} ($\mathcal{T}$) are computed in \texttt{Sailfish} \citep{Westernacher_Schneider_2022}. We present these timeseries data for a selection of different viscosities in Figs.~\ref{fig:AccretionRates},~\ref{fig:PowerandTorque},~\ref{fig:AccretionVsGravitational}.\\

The total accretion rate for low-viscosity disks (top row of Fig.~\ref{fig:AccretionRates}) exhibits irregular structure with strong variability. We note that during the `gentle-sink' relaxation period, the accretion rates are smoother which we attribute to the presence of the minidisks\footnote{Which are only present during the gentle-sink period, $\gamma_\mathrm{sink} = 1\Omega_0$.} which regulate the flow of material around the sink. Once the sink rate is increased and the buffer of the minidisks is lost, interesting behaviour manifests through quasi-periodic accretion `flares' (most visible in the upper right panel of Fig.~\ref{fig:AccretionRates}). Typically, this consists of an initially quiescent period of steady accretion which gradually rises before culminating in a sharp flare (see also Fig.~\ref{fig:AccretionFlare}). These flares soon subside facilitating a return to a steady, quiescent accretion after a few binary orbits. We attribute these flares to phases of instability in the intrabinary bridge, which we discuss further in Section \ref{sec:BridgeInstabilities}. With increasing viscosity comes more regular accretion behaviour, which we attribute to viscosity's role as a damping mechanism.\\

\setcounter{figure}{5}
\begin{figure*}
\centering
    \includegraphics[width=\linewidth]{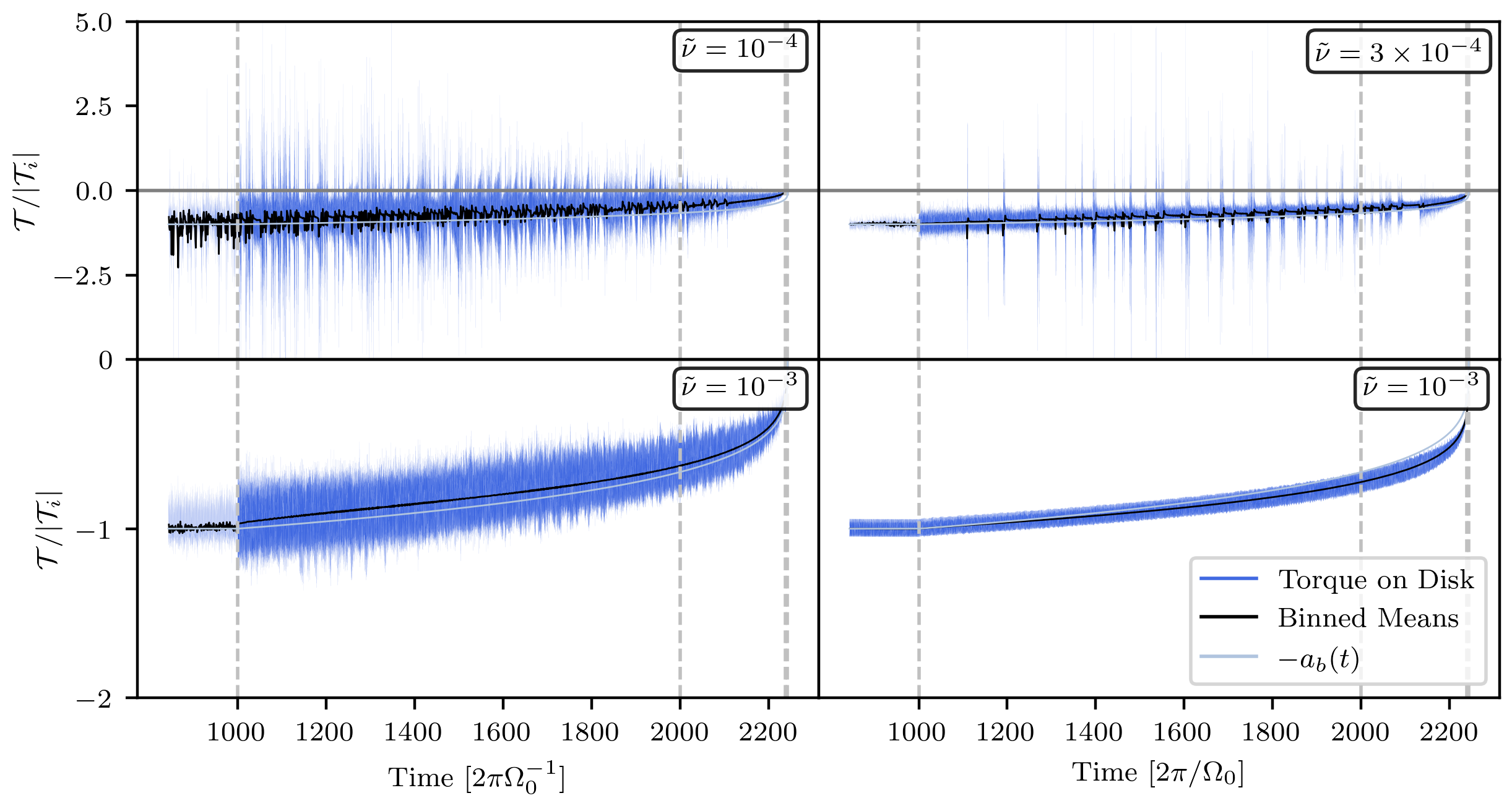}
    \caption{The torque exerted on the disk by the binary (blue) normalised by the magnitude of the steady-state value determined during the burn-in phase $|\mathcal{T}_i|$. A negative torque means that the disk is losing angular momentum to the binary. The black line denotes the binned means of the torque, while the light blue line denotes (minus) the binary semi-major axis $-a_b(t)/a_0$. Similar to the accretion rates, less viscous disks (top) exhibit strong variability in their torque profiles, while larger viscosities (bottom) corresponds to more regular profiles.}
    \label{fig:PowerandTorque}
\end{figure*}

The binary accretion rates steadily decline following the onset of GW-driven inspiral in all panels of Fig.~\ref{fig:AccretionRates}. This secular decline can be attributed to (i) a lack of supply due to inadequate viscous transport of material or (ii) the ``weakening-torque effect'' \citep{zrake2024changinglookinspiralstrendsswitches} where a time-varying torque $\mathcal{T}$ exerted by the binary alters the disk structure, causing long-term trends in the accretion rate. If one considers the relative radial velocity between the binary and an annulus of gas at radius $r$, at some $r$ this relative velocity will be too slow to adequately replenish regions interior to $r$.
That is, before the nominal point of decoupling the binary successively loses viscous contact with each annulus in the disk (from outside in), restricting the viscous inflow of material during inspiral. For example, lower viscosity disks are unable to retain their initial surface density profile during inspiral in Fig.~\ref{fig:AverageSurfaceDensity}, limiting the amount of gas available to be accreted, and slowly `starving' the binary -- even before the nominal point of decoupling. 
In contrast, the larger viscosity $\tilde{\nu} = 10^{-2}$ systems (bottom right panel), retain their initial density profiles throughout the inspiral.
% the disk is capable of replenishing material, retaining its initial density profile throughout the inspiral. 
This suggests that for highly viscous disks, the diminishing accretion rate is not due to the inadequate viscous transport of material, but rather the binary's own inefficiency at accreting gas because of the time-varying torque between the binary and the disk.\\

This weakening torque effect as it manifests here is described by the type-A inspiral models of \cite{zrake2024changinglookinspiralstrendsswitches}, where the disk is in-net delivering angular momentum to the binary. In this scenario, the inner regions of the disk are depleted relative to a disk that is experiencing no external torques (to accommodate the loss of angular momentum and centrifugal support). As the binary loses angular momentum to gravitational waves, the magnitude of the torque sourcing this angular momentum current decreases, and the disk will start to replenish its inner annuli. Mass conservation dictates that while the inner regions accumulate material (at a fixed supply rate), the accretion rate across the disk's inner edge must drop. This attenuates the mass flow to the binary causing the accretion rate to continuously diminish as the strength of GW emission grows. For a retrograde configuration this is consistent with disk-driven orbital decay because the disk delivers angular momentum to the binary, but the binary angular momentum is of opposite sign, so the orbit shrinks. Thus, when the binary evolves due to gravitational radiation, the system is of Type-A and should present with a power-law decay of the accretion rate. This is precisely what is observed, and when we directly compare to the \citet{zrake2024changinglookinspiralstrendsswitches} model, we find that the weakening torque effect accurately describes our accretion rate decay during inspiral (dashed-blue lines in Fig.~\ref{fig:AccretionRates}).\\

In Fig.~\ref{fig:PowerandTorque} the magnitude of the total torque declines approaching the point of merger. The decreasing binary semi-major axis offers a shorter lever arm over which the force is applied, which accurately accounts for the weakening torque in Fig.~\ref{fig:PowerandTorque} (light blue lines). Correspondingly, the force exerted on the binary by the disk remains approximately constant over the duration of the inspiral. We note that for a circular binary, the power $\mathcal{P}$ is proportional to the torque $\mathcal{P} = \Omega \mathcal{T}$ which increases in magnitude approaching the point of merger.

\setcounter{figure}{6}
\begin{figure}[H] 
    \centering
    \hspace{5mm}
    ${\langle{\mathcal{T}_g}\rangle} / \langle{\mathcal{T}_a}\rangle$
    \includegraphics[width=0.97\linewidth]{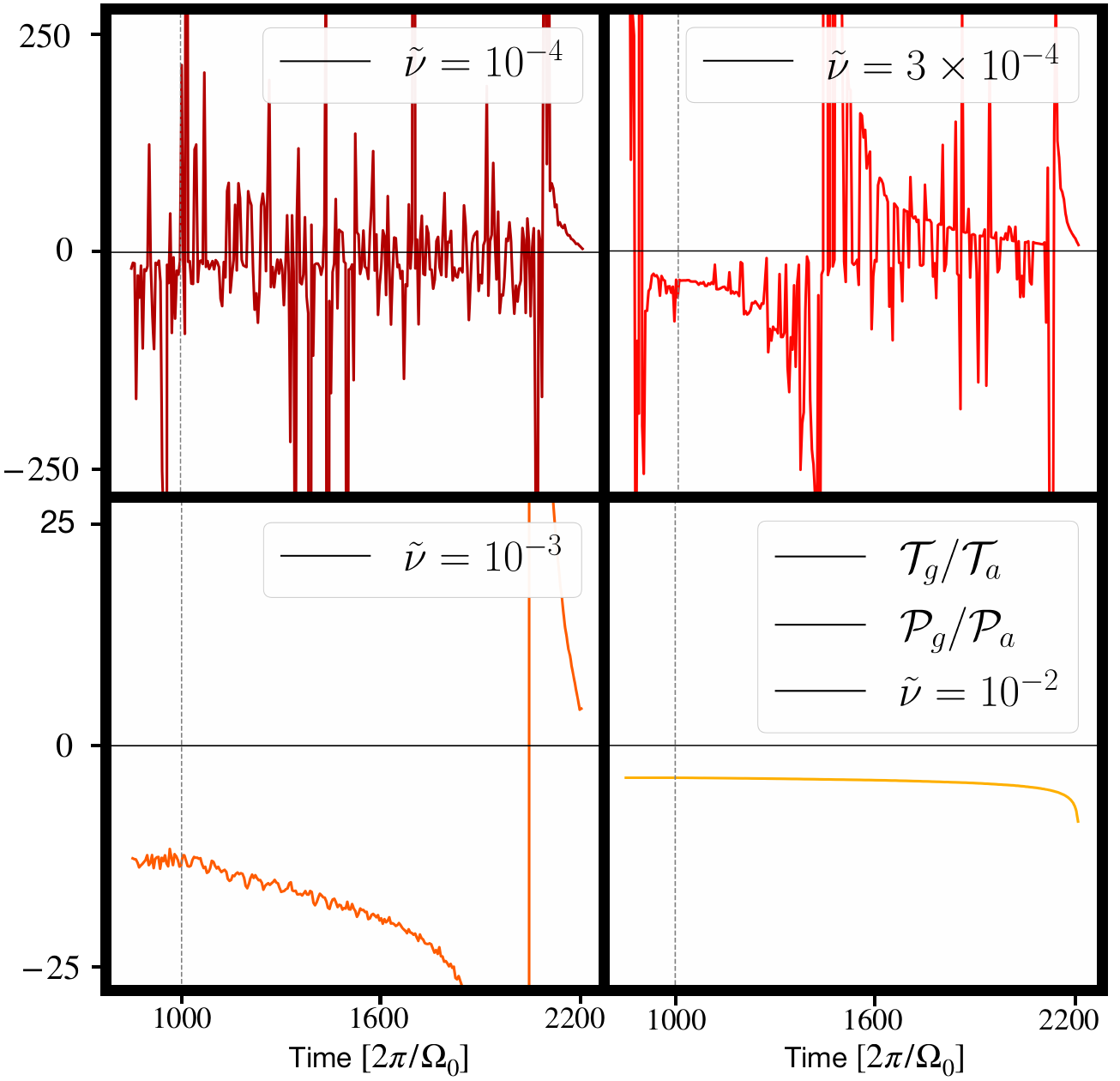}
    \caption{The ratio of gravitational torques to accretion torques as a function of time for different viscosities in each panel.
    }
    \label{fig:AccretionVsGravitational}
\end{figure}

Fig.~\ref{fig:AccretionVsGravitational} illustrates the ratio of the gravitational torque ($\mathcal{T}_g = -\mathbf{a}_b\times{\mathbf{F}_\mathrm{g}}$) to the accretion torque ($\mathcal{T}_a = -\mathbf{a}_b\times{\mathbf{F}_\mathrm{a}}$) for a selection of different viscosities across the inspiral. Gravitational torques generally dominate over accretion torques, with the ratio typically negative. This indicates that the torque from accreted material typically opposes the torque from gravitational forces. This relationship is reflected in Fig.~\ref{fig:BridgeFormation}, where the intrabinary bridge attaches to the rear of the massive body, funneling accreted material into the sink from behind the binary. As this material carries momentum in the direction of motion of the binary, its accretion generates a positive torque, contrasting with the always-negative gravitational torque. Notably, for $\tilde{\nu} = 10^{-3}$ (lower left panel of Fig.~\ref{fig:AccretionVsGravitational}), the ratio between gravitational torque and accretion torque diverges accompanied by a change of sign at $t \approx 2050~ [2\pi/\Omega_0]$. Around this time, the accretion forces have transitioned from being positive (driving) to negative (frictional)\footnote{The change in sign of the accretion force is not visible in Fig.~\ref{fig:PowerandTorque} due to its subdominance to the gravitational force}, signifying that material is no longer being channeled into the sink from behind. As such, the structure of the bridge has changed, suggesting that an instability in the intrabinary bridge has occurred\footnote{The divergence in the torque ratio does not precisely align with the instability in the intrabinary bridge. While most accreted material is funneled through the bridge, some accreted gas is not. Therefore, zero accretion torque does not correspond to the exact moment of bridge instability}. Finally, we note that similar divergences occur at lower viscosities (upper panels of Fig.~\ref{fig:AccretionVsGravitational}) albeit much more frequently. This suggests that bridge instabilities are viscosity dependent phenomenon (discussed further in Section~\ref{sec:BridgeInstabilities}).\\

\subsection{The Decoupling Process}
\label{sec:Decoupling}
We next discuss the process of decoupling for both prograde and retrograde circumbinary disks. In Fig.~\ref{fig:Decoupling}, we plot the cavity semi-major axis as a function of the binary semi-major axis on a logarithmic scale.\\

At early times, the cavity is capable of tracking the binary's inspiral. This is demonstrated by the power law relationship, $a_\mathrm{c} \propto a_\mathrm{b}^p$ for some viscosity dependent $p$. At late times the cavity enters a plateau, conveying the disk's inability to follow the binary. Between these two regimes, there exists an intermediate state where the power law ``breaks'', signifying the decoupling of the binary from the disk. In Fig.~\ref{fig:Decoupling}, the binary can decouple from the disk at a range of binary semi-major axes between $50 r_G \to 5 r_G$ depending on the disk viscosity. We note that our results for prograde decoupling are in good agreement with Fig.~2 of \cite{Dittmann_2023}.\\

Both prograde and retrograde disks exhibit similar decoupling profiles, breaking at comparable semi-major axes in Fig.~\ref{fig:Decoupling}. We thus expect any observational signatures of decoupling to occur at a time set by the viscosity -- not the orbital configuration of the disk. Consequently, time-domain measurements could infer the viscosity by measuring the time of decoupling \citep[the time of decoupling is expected to coincide with a change in the EM signature, \eg~][]{10.1093/mnras/stad3095} without knowing the sense of rotation of the circumbinary disk a priori. Furthermore, the point of decoupling is likely to be within the LISA band \citep[][depending on the black hole masses and disk viscosity]{Dittmann_2023} enabling triggered measurements of the event through multimessenger astronomy. Finally, we note that the break in the power law is smooth rather than sharp, suggesting that decoupling is a gradual process and may be accompanied by long-term EM variability \citep[see][for further discussion]{zrake2024changinglookinspiralstrendsswitches}.\\

The main difference between the two circumbinary disk configurations in Fig.~\ref{fig:Decoupling} is the size of the cavity, which is invariably smaller for retrograde disks. A smaller cavity enables gas to reach deeper into the potential well of the binary, likely leading to the emission of high frequency EM radiation from the circumbinary disk (enhanced UV for example). We note, however, that the lack of minidisks will result in a reduction (or absence) of the highest frequencies of X-ray emission.

\setcounter{figure}{7}
\begin{figure}[H]
    \centering
    \includegraphics[width=\linewidth]{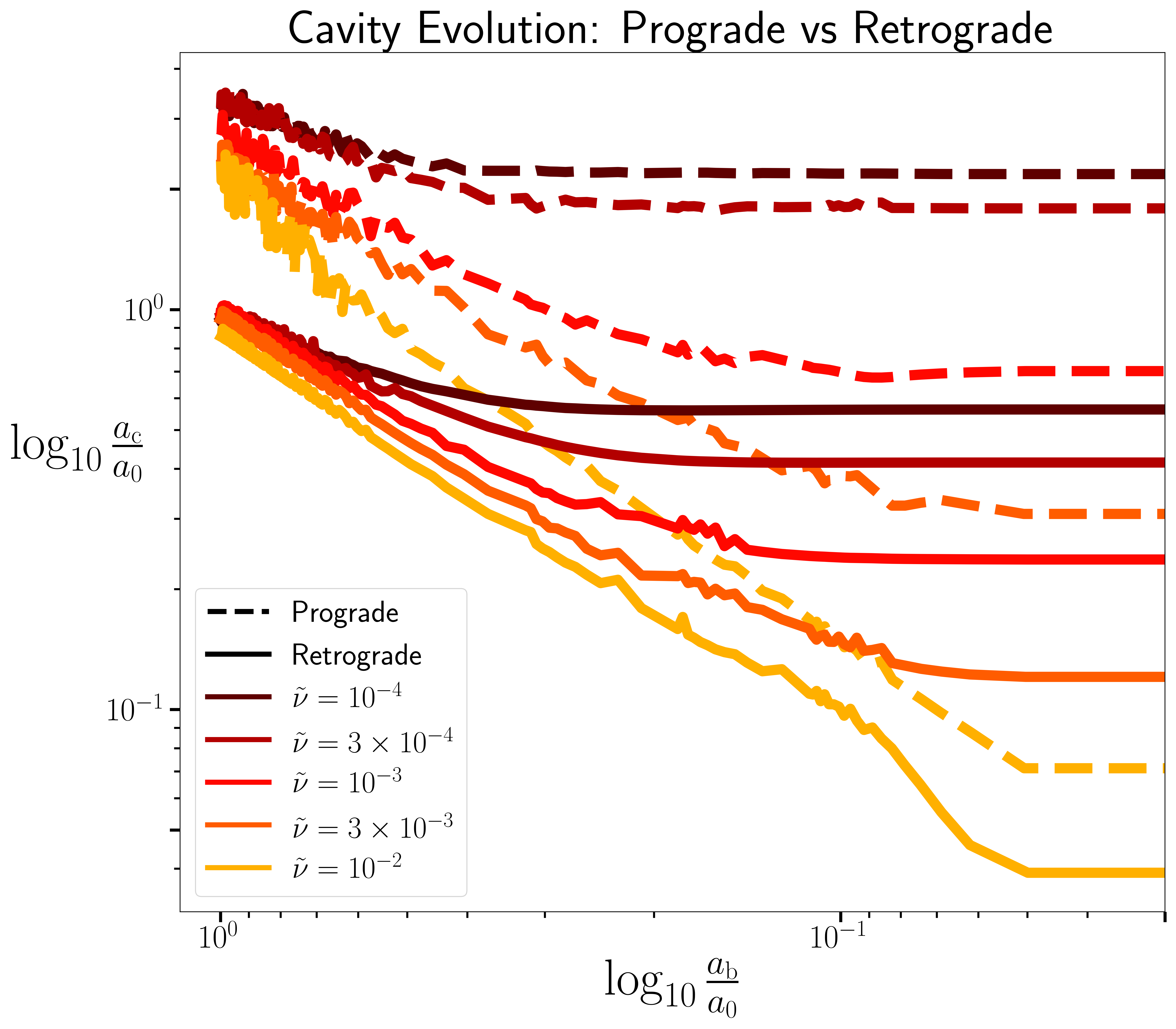}
    \caption{A logarithmic plot of the cavity semi-major axis $a_\mathrm{c}$ as a function of the binary-semi major axis $a_\mathrm{b}$ throughout the inspiral (early times to the left, later times to the right). The dashed and solid lines represent prograde and retrograde disk configurations, respectively, while the different colours represent different kinematic viscosities.}
    \label{fig:Decoupling}
\end{figure}

\section{Discussion}
\subsection{The Stability of Retrograde Orbits}
\label{sec:RetrogradeStability}
Retrograde disks display a small central cavity, suggesting an enhancement in the orbital stability for retrograde circumbinary test particles. In order to gain intuition for the size of the cavity, we review the restricted 3-body problem (R3BP) for disk particles around an equal-mass binary.\\

In the R3BP, we consider a small test particle $m\ll M$ moving with velocity $v$ in the binary's corotating frame of reference. We center our coordinate system on the barycenter from which the test particle is a radial distance $r$ away. In this non-inertial frame, the particle evolves under an effective potential $U_\mathrm{eff}$,\vspace{-1mm}
\begin{equation}
    U_\mathrm{eff} = \frac{m\Omega_\mathrm{b}^2r^2}{2} + \frac{GM}{2r_1} + \frac{GM}{2r_2},
    \label{EffectivePotential}
\end{equation}
where $r_1, r_2$ are the radial distances to the two binary components and $\Omega_\mathrm{b}$ is the binary's orbital frequency. While the total energy and momentum are globally conserved, the only locally conserved quantity (ie. for each test particle individually) is the Jacobi constant $C_J$ defined as,
\begin{equation}
    C_J = 2 U_\mathrm{eff} - v^2.
    \label{JacobiConstant}
\end{equation}
As a constant of motion, $C_J$ is equal at all points along the particle's trajectory. In particular, when evaluated at $v = 0$ the Jacobi constant traces isocontours of the effective potential known as the ``zero velocity curves'' (ZVCs). These ZVCs cannot be crossed, as doing so would require a negative squared velocity. Therefore, these curves delineate the edges of spatial domains accessible to the test particle, and confine it to within the bounds of its ZVC. We illustrate four different examples of ZVCs (corresponding to different values of the Jacobi constant) in Fig.~\ref{fig:ZVCs}. The critical value of the Jacobi constant $C_J^*$ \citep{D_Orazio_2016} encloses the binary (black dots) and separates it from the outer regions of the disk. For subcritical Jacobi constants ($C_J<C_J^*$ red, dashed lines), test particles can cross from the inner to the outer disk through the Lagrange points $L_2$ or $L_3$. Meanwhile, for supercritical Jacobi constants ($C_J\geq C_J^*$, blue dotted lines), test particles cannot cross between the inner and outer disk. Therefore, particles with subcritical Jacobi constants can be expelled by the binary in the process of cavity formation, whereas particles with supercritical Jacobi constants cannot, thus limiting the extent of gap clearing. The isocontour $C_J^R$ (black, dashed line) delineates the binary's Roche-bound region, within which, particles with $C_J > C_J^R$ are bound to an individual component of the binary. This Roche-bound region is an approximation of the maximum size attainable by the circumsingle minidisks.\\

We compute the values of the Jacobi constant for prograde ($-$) and retrograde ($+$) Keplerian disks by prescribing the velocity profile in the corotating frame $v_\mathrm{K,\pm}$,
\begin{equation}
    v_\mathrm{K,\pm} = \Omega_\mathrm{b} r \pm \sqrt{\frac{GM}{r}}.
\end{equation}
In Fig.~\ref{fig:DynamicalExclusion} we segregate Keplerian disks into the regions of subcritical (grey), supercritical (orange) and Roche-bound (beige) Jacobi constant values. This illustrates the disk regions which are more susceptible to being cleared by the prograde binary (grey) and those which are more resilient (orange, beige). While insightful into the dynamics, the Jacobi constant analysis does not provide a measure of orbital
stability other than whether a particle can be expelled
or not. To quantify the stability of orbits within the subcritical regions, we numerically integrate the orbits of three test particles in Fig.~\ref{fig:DynamicalExclusion} using an adaptive time step Runge-Kutta-Fehlberg method, a 4th order integrator with 5th order error estimation. We conserve the Jacobi constant to a fractional error of less than $10^{-4}$ for each particle over the course of the integration ($\sim 3$ binary orbits).

\setcounter{figure}{8}
\begin{figure}[H]
    %\centering
    \hspace{12mm}
    \textbf{Zero Velocity Curves} \\[0.5em]  
    %\hspace{-8mm}
    \includegraphics[width=\linewidth]{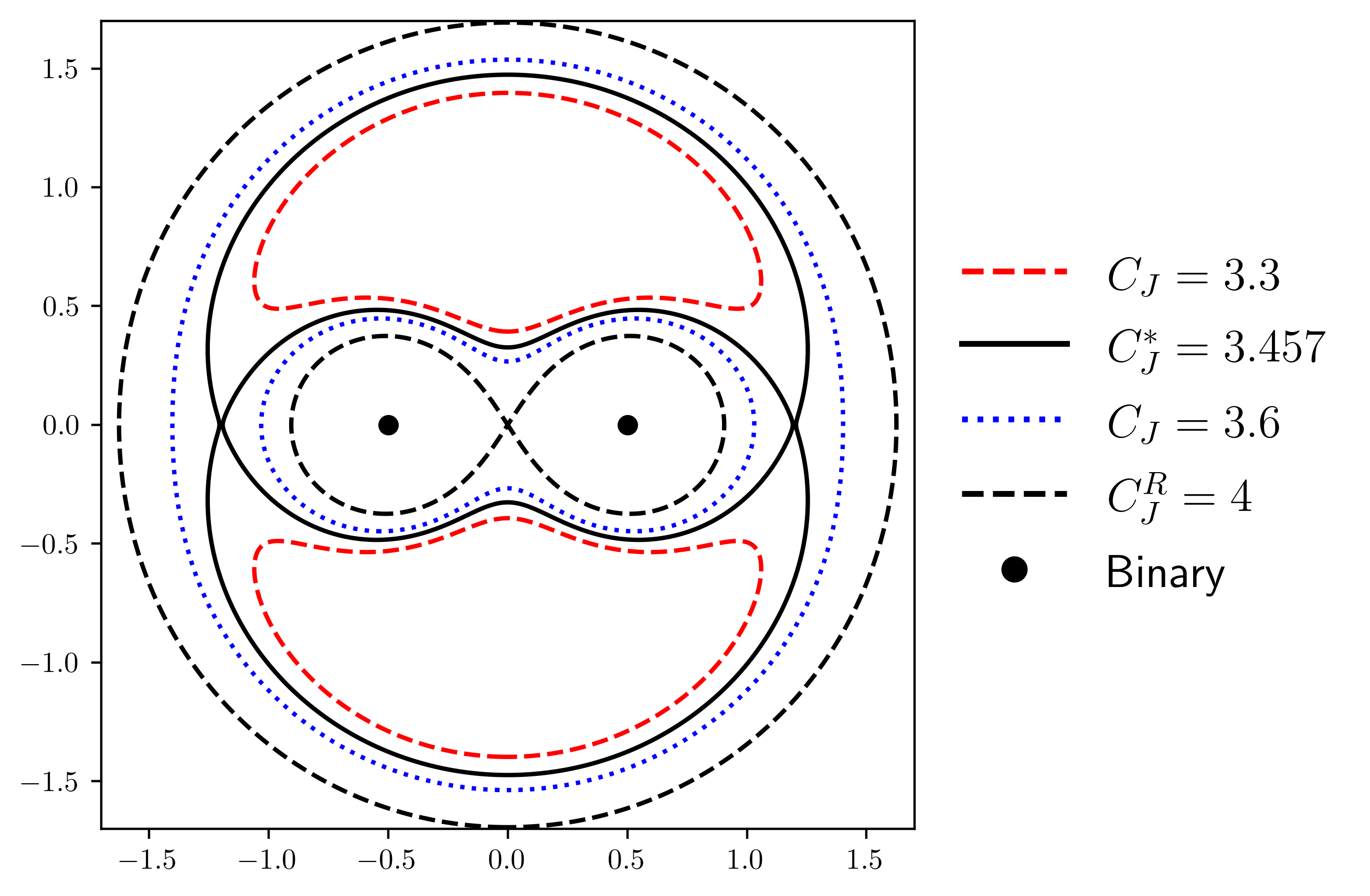}
    \caption{Zero-velocity curves for different values of the Jacobi constant $C_J$. The critical value $C_J^*$ (solid black line) is the minimum value which separates the binary (black dots) from the outer regions of the disk. For $C_J<C_J^*$ (red dashed line), a particle can cross through $L_2$ or $L_3$ to travel between the inner and outer regions of the disk. The Roche-bound region (dashed black line) is described by an isocontour $C_J^R$ and describes test particles that are bound to one of the binary components.}
    \label{fig:ZVCs}
\end{figure}

Retrograde disks have a subcritical region (grey) extending outward from $r \approx 0.7 a_b$ in Fig.~\ref{fig:DynamicalExclusion}. Test particles within this region are stable \citep{10.1111/j.1365-2966.2012.21151.x} almost down to the binary semi-major axis. In contrast, the subcritical region in prograde disks is truncated at radius $r \approx 2 a_b$, wherein, test particle orbits are unstable and can be ejected (see Fig.~\ref{fig:DynamicalExclusion} and Fig.~5 of \citealt{D_Orazio_2016}). In the outer supercritical region (orange), material is unable to access the inner regions of the disk and, therefore, unable to replenish the cavity created by ejected particles. We note that for a \emph{collisional} medium (where the mass of the particles is non-negligible), \cite{MastrobuonoBattisti2024} found that prograde orbits form closer to the center than retrograde ones. Finally, we note that the radius of the Roche-bound regions (beige in Fig.~\ref{fig:DynamicalExclusion}) are $r_{RL} \approx 0.15 a_b$ for retrograde disks and $r_{RL} \approx 0.31 a_b$ for prograde disks, in agreement with the discussion presented in Section \ref{sec:DiskMorphology}.
\setcounter{figure}{9}
\begin{figure}[H]
    \centering
    \includegraphics[width=0.82\linewidth]{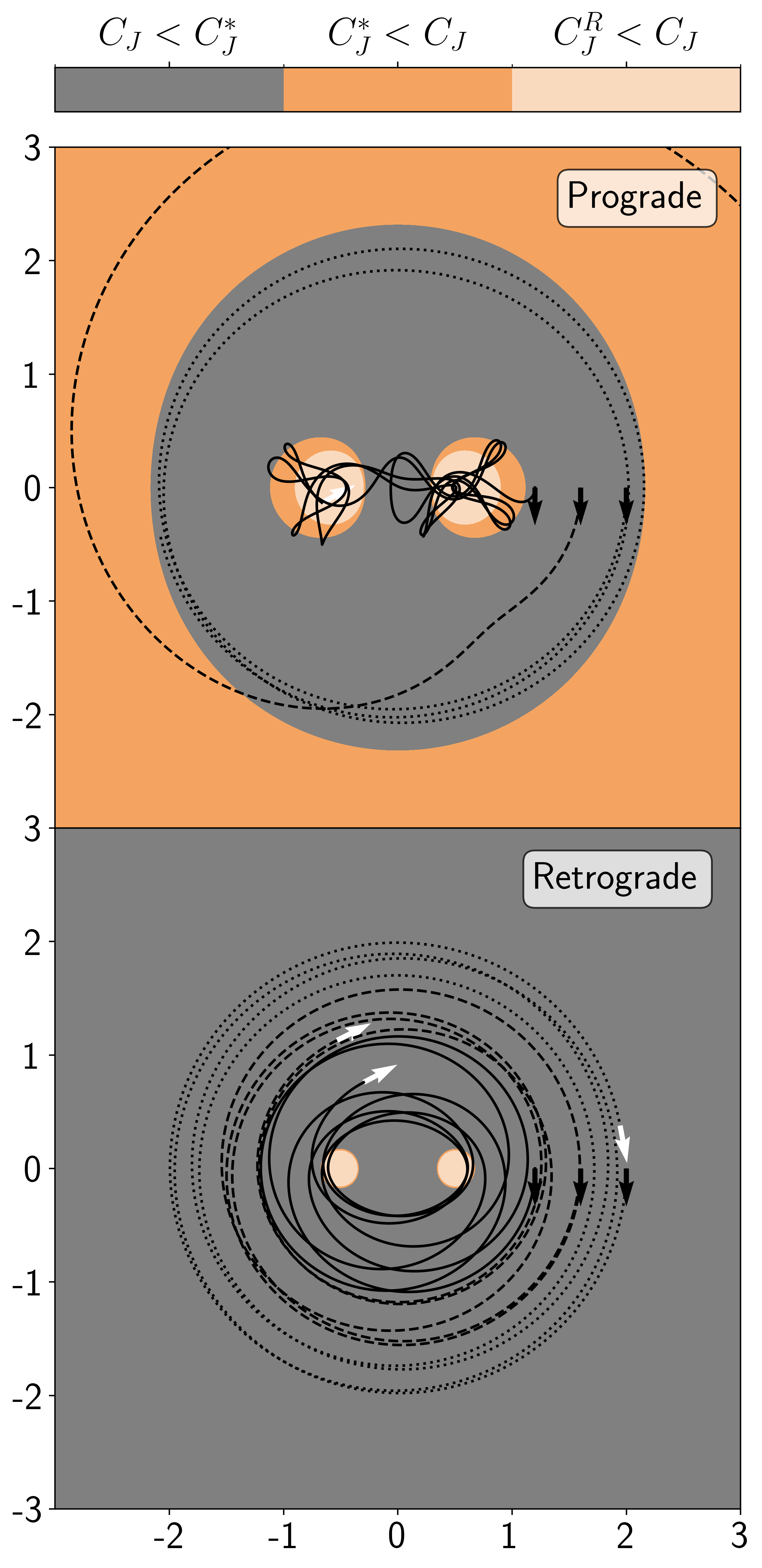}
    \caption{Subcritical (grey), supercritical (orange) and Roche-bound (beige) values of the Jacobi constant for prograde (top) and retrograde (bottom) Keplerian disks. The solid, dashed and dotted lines are tracks of test particles in the corotating frame, initialised on circular Keplerian orbits at radii $r/a_0 = 1.2, 1.6, 2$. The black and white arrows denote the particles initial and final positions respectively.}
    \label{fig:DynamicalExclusion}
\end{figure}
\setcounter{figure}{10}
\begin{figure*}
    \centering
    \includegraphics[width=\linewidth]{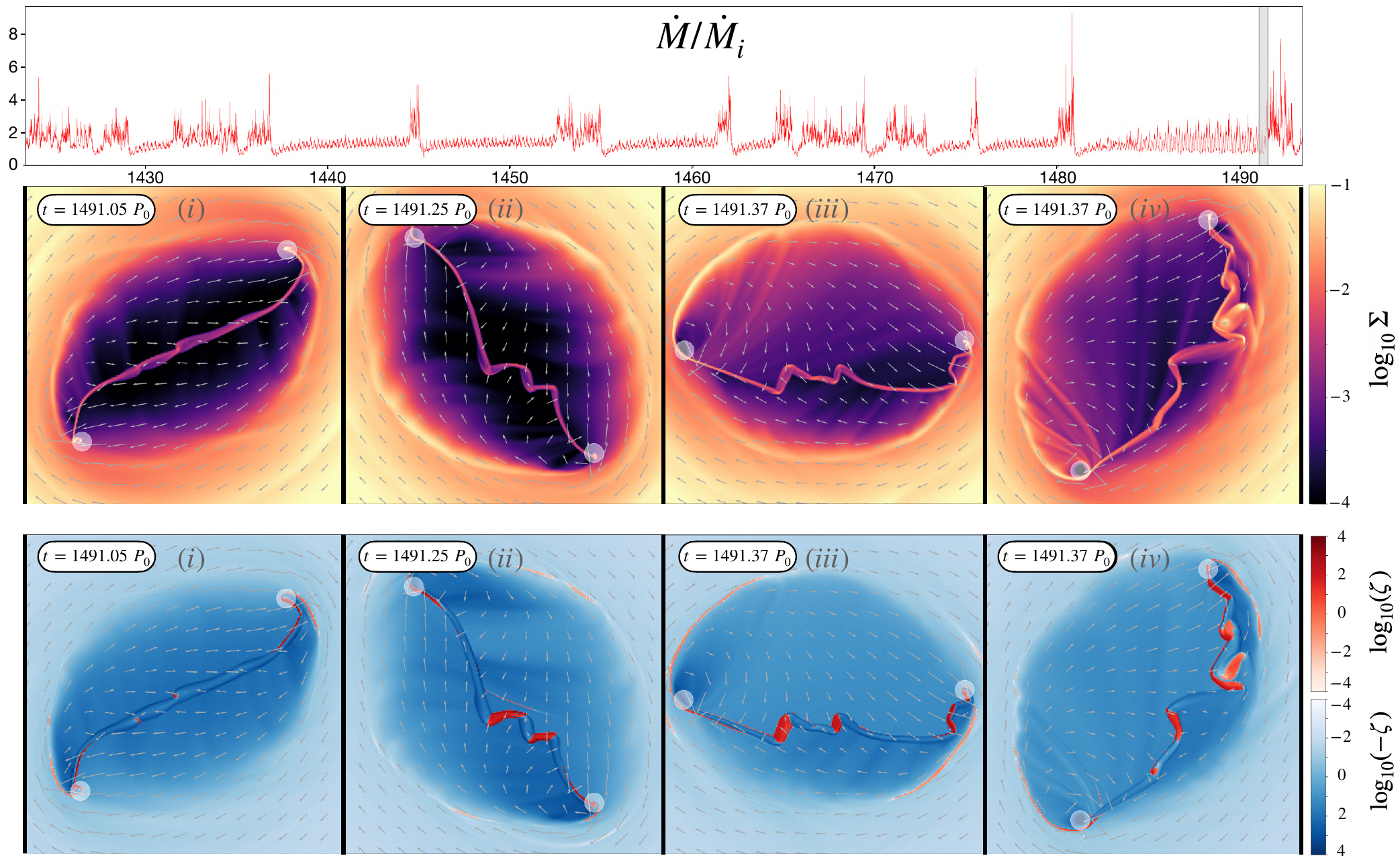}
    \caption{The normalised accretion rate (\emph{top}), disk surface density (\emph{middle}) and vortensity (\emph{bottom}) during a phase of bridge instability (grey band) for $\tilde{\nu}= 10^{-4}$. The arrows illustrate the velocity profile of the disk, while the white circles represent the component sinks. The vortensity can be either aligned (red) or misaligned (blue) with the binary angular momentum vector. The white vectors illustrate the local gas velocity while the binary moves counter clockwise. The time in each panel is expressed in units of the initial orbital period $P_0 = 2\pi\Omega_0^{-1}$.
    }
    \label{fig:AccretionFlare}
\end{figure*}
\subsection{Bridge Instabilities}
\label{sec:BridgeInstabilities}
Low-viscosity retrograde disks exhibit quasi-periodic flaring, corresponding to phases in which the accretion rate and torque sharply rise. Coincident with these flares are dynamical instabilities in the intrabinary bridge, where it dislodges and re-stabilises a few orbits later. We illustrate an example of this instability in Fig.~\ref{fig:AccretionFlare}.
\pagebreak

The intrabinary bridge serves as an interface between two regions of the disk, where the direction of flow reverses. This reversal results in a steep velocity gradient, making the bridge prone to Kelvin-Helmholtz (KH)-like instabilities. We propose that the vortices generated by the KH mechanism act as a driving source for this instability. To capture the vortex structure along the intrabinary bridge, we illustrate the disk surface density and vortensity ($\zeta = \Sigma^{-1}\left(\nabla\times\mathbf{v}\right)$) in Fig.~\ref{fig:AccretionFlare} over a series of times during the instability.\\

In column $(i)$, the bridge is in a stable configuration with a steady accretion rate (grey band, top row of Fig.~\ref{fig:AccretionFlare}). In column $(ii)$, vortices emerge along the intrabinary bridge (bottom row), causing it to \emph{knot} and \emph{twist} as they propagate toward the sinks. By column $(iii)$, these vortex-induced kinks have displaced the bridge from its natural equilibrium, allowing it to lopsidedly sway out of place. In column $(iv)$, the sway of the bridge becomes more pronounced as it moves further from equilibrium, colliding with the cavity wall and disrupting the flow of gas in the circumbinary disk. In this column, the bridge instability is apparent, characterised by complicated gas dynamics and a dislodged intrabinary bridge. Similarly, the instability is reflected in the accretion timeseries by a flare occurring at $t\approx 1491.5~[2\pi\Omega_0^{-1}]$. This behaviour only manifests at low viscosities as vortex generation is surpressed for more viscous disks, and as a result, the bridge remains more stable. We note that \cite{Tiede_2023} did not see this behaviour due to a higher kinematic viscosity ($\tilde{\nu} = 10^{-3}$).\\

We find little to no sensitivity on the inner boundary sinks, with the instability persisting for (a) different sink prescriptions (both acceleration free and torque free, cf. \citealt{DittmannRyan2021}) and (b) lower sink rates (for $\gamma_\mathrm{sink} = 1$ and $50$). Additionally, the instability persists for higher spatial resolution (increased to $n = 6000 \times 6000$ over the same domain). The instability even occurs during the gentle sink period, meaning that when the sink is surrounded (and buffered) by the minidisk, the bridge can still become unstable affirming that the sink has little role in sourcing the dynamics. In addition, the gentle sink period is at a time before the inspiral has begun, ruling out any effects from asymmetries associated with the triangular Lagrange points $L_4$ and $L_5$ drifting from their fixed positions \citep{Schnittman_2010}. Therefore, it appears that the accretion flares and bridge instabilities are sourced by underlying physical processes rather than numerical artifacts and are unrelated to the inspiral.\\

Although the bridge instability is not sourced by the binary inspiral, it becomes more extreme at later times. During these late times, the intrabinary bridge can never stabilise (for low viscosities), evident from flares in Fig.~\ref{fig:AccretionRates} and visible in Fig.~\ref{fig:DiskMorphology}. Without the typical cycle into and out of equilibrium, the accretion rate remains in a constant state of large variability.

\subsection{Observational Appearance}
Gravitational wave inspirals in prograde and retrograde disks share several similarities while also exhibiting significant variations. We briefly discuss the prospect of using EM observations to distinguish between the two disk configurations.\\ 

\begin{figure}[H]
    %\centering
    \hspace{-5mm}
    \includegraphics[width=\linewidth]{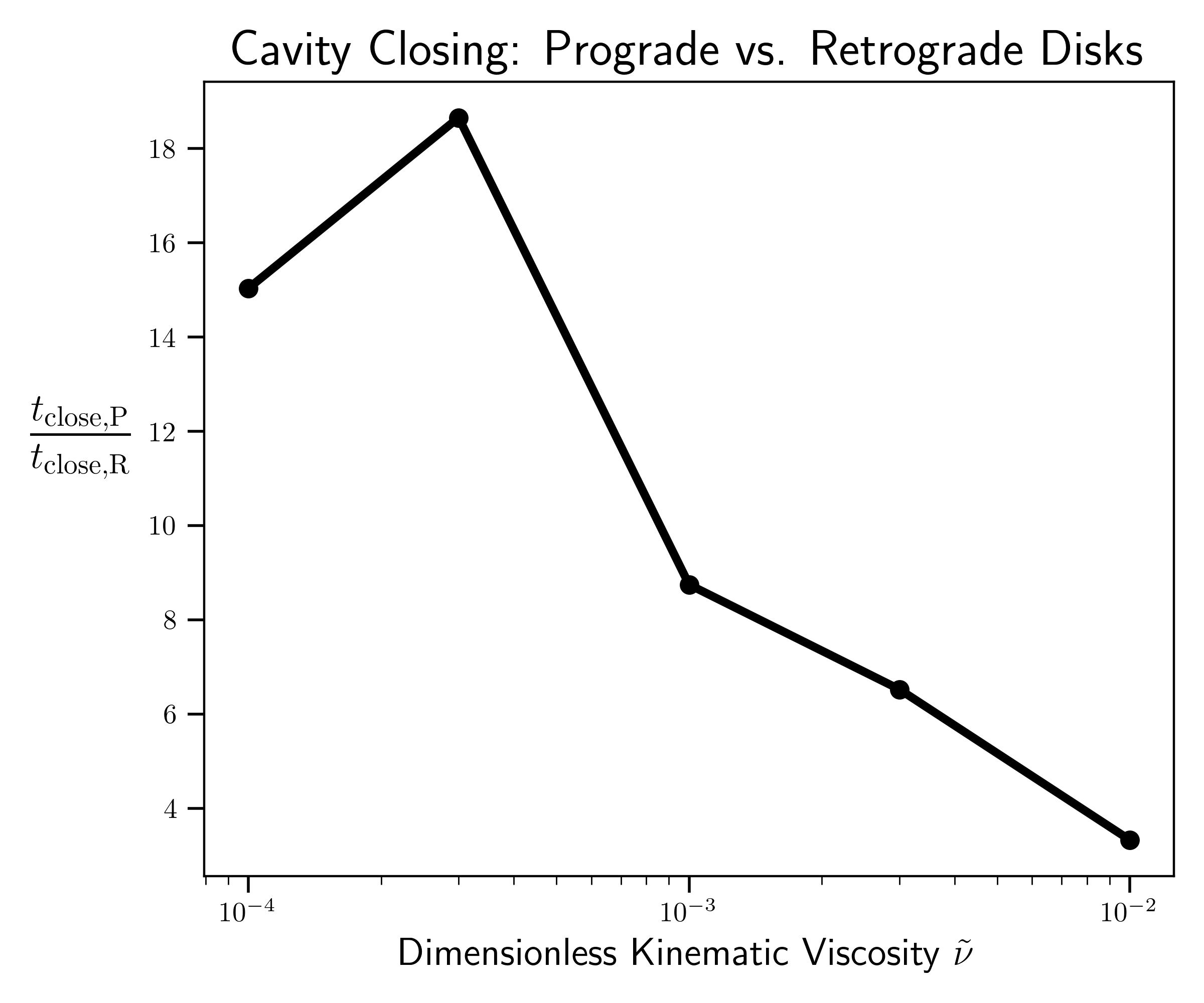}
    \caption{The ratio of prograde to retrograde cavity closing timescales (or disk rebrightening), following merger. These timescales are based on the cavity semi-major axes at decoupling (given in Fig.~\ref{fig:Decoupling}).}
    \label{fig:ClosingTimescale}
\end{figure}

While decoupling occurs at nearly equal times between prograde and retrograde disks (for equal viscosities), the cavity plateau is noticeably lower for retrograde disks (see Fig.~\ref{fig:Decoupling}). This will lead to retrograde disks having:
\begin{itemize}
    \item Higher frequency EM emission from the circumbinary disk at the point of decoupling (and prior), likely resulting in increased optical and UV luminosities,
    \item A shorter cavity closing timescale, or equivalently, a shorter ``rebrightening'' timescale after merger.
\end{itemize}
As a simple analytical estimate of the re-brightening timescale, we compute the cavity's viscous time $t_\mathrm{close} = a_\mathrm{c}^2 / 3\nu$ from the semi-major axis it had at decoupling in Fig.~\ref{fig:ClosingTimescale}.\\

The eccentric, lopsided nature of prograde cavities will enable some material to reach the merged remnant before $\tau_\nu$ in Fig.~\ref{fig:ClosingTimescale}. Nonetheless, a \emph{full rebrightening} will require the entire cavity to have closed, possibly taking an order of magnitude longer to occur for a prograde disk than a retrograde one, possibly providing a way to distinguish between both disk configurations. The ratio of rebrightening timescales is not monotonic with viscosity, peaking around $\nu = 3\times 10^{-3}a_0^2\Omega_0$ and declining for larger viscosities -- suggesting that the size of prograde and retrograde cavities do not scale equally with viscosity. In addition, retrograde merger remnants will have intrinsic spin oriented oppositely to the rotation of the refilling disk \citep[\eg~][]{Mudit2024}, resulting in a larger ISCO than prograde mergers. The refuelling ISCO is potentially observable via FeK-$\alpha$ line measurements if sufficiently bright, providing another way to distinguish between the two disk configurations.\\

The appearance of the intrabinary bridge may be a distinguishing feature of retrograde circumbinary disks. The shocked gas will likely produce the highest frequencies of EM radiation, which could be lensed by the binary, and Doppler boosted by its dynamics \citep{DOrazio2017, DOrazio2024}. In addition, the quasi-periodic flaring associated with bridge instabilities may be a general feature of low-viscosity retrograde disks and could appear as repeating nuclear transients characterised by X-ray flaring. Understanding the exact observational signatures produced will require non-isothermal simulations \citep{Tanaka_BirthofaQuasar} and is a topic left for future work.\\ 

\section{Summary}
We have performed numerical hydrodynamic simulations of retrograde circumbinary disks, hosting an equal-mass binary undergoing gravitational wave inspiral. We have compared our results directly with prograde counterparts, to better understand the differences between prograde and retrograde disk configurations. We found that:
\begin{itemize}
    \item Both prograde and retrograde disks decouple from the binary at comparable binary semi-major axes (for equal viscosities). 
    \item Retrograde central cavities are significantly smaller than prograde ones, explained by the presence of stable orbits close to the binary. This will lead to the circumbinary disk producing higher frequency emission at the point of decoupling. Additionally, this will cause the cavity to close earlier post-merger, leading to a faster rebrightening of the disk.
    \item Low-viscosity retrograde disks are prone to bridge instabilities, leading to quasi-periodic accretion flares. These may produce distinctive EM signatures of binaries in retrograde disks.
    \item The lack of X-ray luminous minidisks in the retrograde case will reduce the amount of high-frequency EM emission, leading to different emission spectra than prograde counterparts. 
    In particular, because the disruption of prograde minidisks may signify a temporal signature of merging SMBHBs \citep[\eg, an X-ray turnoff][]{10.1093/mnras/stad3095}, their absence for retrograde disks will alter such time-domain observables.\\

    \item The lack of X-ray luminous minidisks in the retrograde case will reduce the amount of high-frequency EM emission, leading to different emission spectra than prograde counterparts. 
    In particular, the lack of minidisks will likely contrast with the time-domain observables expected for prograde disks \citep[\eg, an X-ray turnoff][]{10.1093/mnras/stad3095}.\\
\end{itemize}

We highlight a number of simplifying assumptions made in this study. The prescribed binary dynamics in Eqs.~(\ref{SemiMajorAxisInspiral}) and (\ref{PhaseChange}) are an approximation to the full general relativistic equations of motion. In addition, the relativistic effects of gas dynamics have been neglected \citep{2022ApJ...928..137G, 2023arXiv230518538A}, along with magnetic fields \citep{2012ApJ...755...51N, most2024decouplingsupermassiveblackhole}, radiation \citep{DelValle2018, Williamson2022}, jets and gravitational wave recoil kicks \citep{10.1093/mnras/stad3095}. Each of these phenomena will likely produce different observable consequences. More realistic disks may also have Mach numbers significantly higher than $\mathcal{M} = 10$, as adopted in this study, likely altering the disk morphology and accretion rate variability \citep{2020ApJ...900...43T,2024arXiv241003830T}. Finally, we note that highly eccentric binaries embedded in massive retrograde disks have been observed to experience tilting instabilities \citep{10.1093/mnras/stu194, Mudit2024} in which the binary can re-orient itself into a prograde configuration. Addressing these caveats is left for future work.

\acknowledgments
DJD, CT, and DON acknowledge support from the Danish Independent Research Fund through Sapere Aude Starting Grant No. 121587, led by DJD.  
This work was also supported in part by the LISA Preparatory Science Program (LPS) through NASA grant 80NSSC24K0440, by NASA Astrophysics Theory Program (ATP) grant 80NSSC22K0822, and by the European Union’s Horizon research and innovation program under Marie Sklodowska-Curie grant agreement No. 101148364.
This work made use of the following software packages: \texttt{Sailfish} \citep{2024ascl.soft08004Z}, \texttt{numpy} \citep{numpy}, \texttt{python} \citep{python}, and \texttt{scipy} \citep{2020SciPy-NMeth, scipy_13352243}. Software citation information aggregated using \texttt{\href{https://www.tomwagg.com/software-citation-station/}{The Software Citation Station}} \citep{software-citation-station-paper, software-citation-station-zenodo}. 
The Tycho supercomputer hosted at the SCIENCE HPC center at the University of Copenhagen was used in this work.

\bibliography{bibliography.bib}

\begin{thebibliography}{}
\expandafter\ifx\csname natexlab\endcsname\relax\def\natexlab#1{#1}\fi
\providecommand{\url}[1]{\href{#1}{#1}}
\providecommand{\dodoi}[1]{doi:~\href{http://doi.org/#1}{\nolinkurl{#1}}}
\providecommand{\doeprint}[1]{\href{http://ascl.net/#1}{\nolinkurl{http://ascl.net/#1}}}
\providecommand{\doarXiv}[1]{\href{https://arxiv.org/abs/#1}{\nolinkurl{https://arxiv.org/abs/#1}}}

\bibitem[{Afroz \& Mukherjee(2024)}]{afroz2024modelindependentprecisiontestgeneral}
Afroz, S., \& Mukherjee, S. 2024, A Model-Independent Precision Test of General Relativity using LISA Bright Standard Sirens.
\newblock \doarXiv{2406.08791}

\bibitem[{{Amaro-Seoane} {et~al.}(2023){Amaro-Seoane}, {Andrews}, {Arca Sedda}, {Askar}, {Baghi}, {Balasov}, {Bartos}, {Bavera}, {Bellovary}, {Berry}, {Berti}, {Bianchi}, {Blecha}, {Blondin}, {Bogdanovi{\'c}}, {Boissier}, {Bonetti}, {Bonoli}, {Bortolas}, {Breivik}, {Capelo}, {Caramete}, {Cattorini}, {Charisi}, {Chaty}, {Chen}, {Chru{\'s}li{\'n}ska}, {Chua}, {Church}, {Colpi}, {D'Orazio}, {Danielski}, {Davies}, {Dayal}, {De Rosa}, {Derdzinski}, {Destounis}, {Dotti}, {Du{\c{t}}an}, {Dvorkin}, {Fabj}, {Foglizzo}, {Ford}, {Fouvry}, {Franchini}, {Fragos}, {Fryer}, {Gaspari}, {Gerosa}, {Graziani}, {Groot}, {Habouzit}, {Haggard}, {Haiman}, {Han}, {Istrate}, {Johansson}, {Khan}, {Kimpson}, {Kokkotas}, {Kong}, {Korol}, {Kremer}, {Kupfer}, {Lamberts}, {Larson}, {Lau}, {Liu}, {Lloyd-Ronning}, {Lodato}, {Lupi}, {Ma}, {Maccarone}, {Mandel}, {Mangiagli}, {Mapelli}, {Mathis}, {Mayer}, {McGee}, {McKernan}, {Miller}, {Mota}, {Mumpower}, {Nasim}, {Nelemans}, {Noble}, {Pacucci}, {Panessa}, {Paschalidis}, {Pfister}, {Porquet},
  {Quenby}, {Ricarte}, {R{\"o}pke}, {Regan}, {Rosswog}, {Ruiter}, {Ruiz}, {Runnoe}, {Schneider}, {Schnittman}, {Secunda}, {Sesana}, {Seto}, {Shao}, {Shapiro}, {Sopuerta}, {Stone}, {Suvorov}, {Tamanini}, {Tamfal}, {Tauris}, {Temmink}, {Tomsick}, {Toonen}, {Torres-Orjuela}, {Toscani}, {Tsokaros}, {Unal}, {V{\'a}zquez-Aceves}, {Valiante}, {van Putten}, {van Roestel}, {Vignali}, {Volonteri}, {Wu}, {Younsi}, {Yu}, {Zane}, {Zwick}, {Antonini}, {Baibhav}, {Barausse}, {Bonilla Rivera}, {Branchesi}, {Branduardi-Raymont}, {Burdge}, {Chakraborty}, {Cuadra}, {Dage}, {Davis}, {de Mink}, {Decarli}, {Doneva}, {Escoffier}, {Gandhi}, {Haardt}, {Lousto}, {Nissanke}, {Nordhaus}, {O'Shaughnessy}, {Portegies Zwart}, {Pound}, {Schussler}, {Sergijenko}, {Spallicci}, {Vernieri}, \& {Vigna-G{\'o}mez}}]{2023LRR....26....2A}
{Amaro-Seoane}, P., {Andrews}, J., {Arca Sedda}, M., {et~al.} 2023, Living Reviews in Relativity, 26, 2, \dodoi{10.1007/s41114-022-00041-y}

\bibitem[{Amaro-Seoane {et~al.}(2023)Amaro-Seoane, Andrews, Arca~Sedda, Askar, Baghi, Balasov, Bartos, Bavera, Bellovary, Berry, Berti, Bianchi, Blecha, Blondin, Bogdanović, Boissier, Bonetti, Bonoli, Bortolas, Breivik, Capelo, Caramete, Cattorini, Charisi, Chaty, Chen, Chruślińska, Chua, Church, Colpi, D’Orazio, Danielski, Davies, Dayal, De~Rosa, Derdzinski, Destounis, Dotti, Duţan, Dvorkin, Fabj, Foglizzo, Ford, Fouvry, Franchini, Fragos, Fryer, Gaspari, Gerosa, Graziani, Groot, Habouzit, Haggard, Haiman, Han, Istrate, Johansson, Khan, Kimpson, Kokkotas, Kong, Korol, Kremer, Kupfer, Lamberts, Larson, Lau, Liu, Lloyd-Ronning, Lodato, Lupi, Ma, Maccarone, Mandel, Mangiagli, Mapelli, Mathis, Mayer, McGee, McKernan, Miller, Mota, Mumpower, Nasim, Nelemans, Noble, Pacucci, Panessa, Paschalidis, Pfister, Porquet, Quenby, Ricarte, Röpke, Regan, Rosswog, Ruiter, Ruiz, Runnoe, Schneider, Schnittman, Secunda, Sesana, Seto, Shao, Shapiro, Sopuerta, Stone, Suvorov, Tamanini, Tamfal, Tauris, Temmink, Tomsick,
  Toonen, Torres-Orjuela, Toscani, Tsokaros, Unal, Vázquez-Aceves, Valiante, van Putten, van Roestel, Vignali, Volonteri, Wu, Younsi, Yu, Zane, Zwick, Antonini, Baibhav, Barausse, Bonilla~Rivera, Branchesi, Branduardi-Raymont, Burdge, Chakraborty, Cuadra, Dage, Davis, de~Mink, Decarli, Doneva, Escoffier, Gandhi, Haardt, Lousto, Nissanke, Nordhaus, O’Shaughnessy, Portegies~Zwart, Pound, Schussler, Sergijenko, Spallicci, Vernieri, \& Vigna-Gómez}]{Amaro_Seoane_2023}
Amaro-Seoane, P., Andrews, J., Arca~Sedda, M., {et~al.} 2023, Living Reviews in Relativity, 26, \dodoi{10.1007/s41114-022-00041-y}

\bibitem[{Armitage \& Natarajan(2002)}]{Armitage2002AccretionDT}
Armitage, P.~J., \& Natarajan, P. 2002, The Astrophysical Journal Letters, 567, L9 .
\newblock \url{https://api.semanticscholar.org/CorpusID:17909303}

\bibitem[{Artymowicz \& Lubow(1996)}]{Artymowicz_1996}
Artymowicz, P., \& Lubow, S.~H. 1996, The Astrophysical Journal, 467, L77, \dodoi{10.1086/310200}

\bibitem[{Auclair {et~al.}(2023)Auclair, Bacon, Baker, Barreiro, Bartolo, Belgacem, Bellomo, Ben-Dayan, Bertacca, Besançon, Blanco-Pillado, Blas, Boileau, Calcagni, Caldwell, Caprini, Carbone, Chang, Chen, Christensen, Clesse, Comelli, Congedo, Contaldi, Crisostomi, Croon, Cui, Cusin, Cutting, Dalang, Luca, del Pozzo, Desjacques, Dimastrogiovanni, Dorsch, Ezquiaga, Fasiello, Figueroa, Flauger, Franciolini, Frusciante, Fumagalli, Garc{\'i}a-Bellido, Gould, Holz, Iacconi, Jain, Jenkins, Jinno, Joana, Karnesis, Konstandin, Koyama, Kozaczuk, Kuroyanagi, Laghi, Lewicki, Lombriser, Madge, Maggiore, Malhotra, Mancarella, Mandic, Mangiagli, Matarrese, Mazumdar, Mukherjee, Musco, Nardini, No, Papanikolaou, Peloso, Pieroni, Pilo, Raccanelli, Renaux-Petel, Renzini, Ricciardone, Riotto, Romano, Rollo, Pol, Morales, Sakellariadou, Saltas, Scalisi, Schmitz, Schwaller, Sergijenko, Servant, Simakachorn, del Sorbo, Sousa, Speri, Steer, Tamanini, Tasinato, Torrado, Unal, Vennin, Vernieri, Vernizzi, Volonteri, Wachter, Wands,
  Witkowski, Zumalac'arregui, Annis, Ares, Avelino, Avgoustidis, Barausse, Bonilla, Bonvin, Bosso, Calabrese, cCalicskan, Cembranos, Chala, Chernoff, Clough, Criswell, Das, Silva, Dayal, Domcke, Durrer, Easther, Escoffier, Ferrans, Fryer, Gair, Gordon, Hendry, Hindmarsh, Hooper, Kajfasz, Kopp, Koushiappas, Kumar, Kunz, Lagos, Lilley, Lizarraga, Lobo, Maleknejad, Martins, Meerburg, Meyer, Mimoso, Nesseris, Nunes, Oikonomou, Orlando, Ozsoy, Pacucci, Palmese, Petiteau, Pinol, Zwart, Pratten, Prokopec, Quenby, Rastgoo, van~der Roest, Rummukainen, Schimd, Secroun, Sopuerta, Tereno, Tolley, Urrestilla, Vagenas, van~de Vis, van~de Weygaert, Wardell, Weir, White, 'Swie.zewska, \& Zhdanov}]{Auclair2023CosmologyWT}
Auclair, P., Bacon, D., Baker, T., {et~al.} 2023, Living Reviews in Relativity, 26, 1.
\newblock \url{https://api.semanticscholar.org/CorpusID:248118657}

\bibitem[{{Avara} {et~al.}(2023){Avara}, {Krolik}, {Campanelli}, {Noble}, {Bowen}, \& {Ryu}}]{2023arXiv230518538A}
{Avara}, M.~J., {Krolik}, J.~H., {Campanelli}, M., {et~al.} 2023, arXiv e-prints, arXiv:2305.18538, \dodoi{10.48550/arXiv.2305.18538}

\bibitem[{Bankert {et~al.}(2015)Bankert, Krolik, \& Shi}]{Bankert_2015}
Bankert, J., Krolik, J.~H., \& Shi, J. 2015, The Astrophysical Journal, 801, 114, \dodoi{10.1088/0004-637X/801/2/114}

\bibitem[{{Barnes} \& {Hernquist}(1991)}]{1991ApJ...370L..65B}
{Barnes}, J.~E., \& {Hernquist}, L.~E. 1991, \apjl, 370, L65, \dodoi{10.1086/185978}

\bibitem[{{Begelman} {et~al.}(1980){Begelman}, {Blandford}, \& {Rees}}]{1980Natur.287..307B}
{Begelman}, M.~C., {Blandford}, R.~D., \& {Rees}, M.~J. 1980, \nat, 287, 307, \dodoi{10.1038/287307a0}

\bibitem[{{Berczik} {et~al.}(2006){Berczik}, {Merritt}, {Spurzem}, \& {Bischof}}]{2006ApJ...642L..21B}
{Berczik}, P., {Merritt}, D., {Spurzem}, R., \& {Bischof}, H.-P. 2006, \apjl, 642, L21, \dodoi{10.1086/504426}

\bibitem[{{Berentzen} {et~al.}(2009){Berentzen}, {Preto}, {Berczik}, {Merritt}, \& {Spurzem}}]{2009ApJ...695..455B}
{Berentzen}, I., {Preto}, M., {Berczik}, P., {Merritt}, D., \& {Spurzem}, R. 2009, \apj, 695, 455, \dodoi{10.1088/0004-637X/695/1/455}

\bibitem[{Bogdanović {et~al.}(2022)Bogdanović, Miller, \& Blecha}]{Bogdanovi__2022}
Bogdanović, T., Miller, M.~C., \& Blecha, L. 2022, Living Reviews in Relativity, 25, \dodoi{10.1007/s41114-022-00037-8}

\bibitem[{Bowen {et~al.}(2018)Bowen, Mewes, Campanelli, Noble, Krolik, \& Zilhão}]{Bowen_2018}
Bowen, D.~B., Mewes, V., Campanelli, M., {et~al.} 2018, The Astrophysical Journal Letters, 853, L17, \dodoi{10.3847/2041-8213/aaa756}

\bibitem[{Casura {et~al.}(2024)Casura, Ilić, Targaczewski, Rakić, \& Liske}]{casura2024exploringmassmeasurementssupermassive}
Casura, S., Ilić, D., Targaczewski, J., Rakić, N., \& Liske, J. 2024, Exploring mass measurements of supermassive black holes in AGN using GAMA photometry and spectroscopy.
\newblock \doarXiv{2408.17275}

\bibitem[{{Chandrasekhar}(1943)}]{Chandrasekhar_DF}
{Chandrasekhar}, S. 1943, \apj, 97, 255, \dodoi{10.1086/144517}

\bibitem[{{Clyburn} \& {Zrake}(2024)}]{2024arXiv240510281C}
{Clyburn}, M., \& {Zrake}, J. 2024, arXiv e-prints, arXiv:2405.10281, \dodoi{10.48550/arXiv.2405.10281}

\bibitem[{Cuadra {et~al.}(2009)Cuadra, Armitage, Alexander, \& Begelman}]{10.1111/j.1365-2966.2008.14147.x}
Cuadra, J., Armitage, P.~J., Alexander, R.~D., \& Begelman, M.~C. 2009, Monthly Notices of the Royal Astronomical Society, 393, 1423, \dodoi{10.1111/j.1365-2966.2008.14147.x}

\bibitem[{{del Valle} \& {Volonteri}(2018)}]{DelValle2018}
{del Valle}, L., \& {Volonteri}, M. 2018, \mnras, 480, 439, \dodoi{10.1093/mnras/sty1815}

\bibitem[{{DeLaurentiis} {et~al.}(2024){DeLaurentiis}, {Haiman}, {Westernacher-Schneider}, {Major Krauth}, {Davelaar}, {Zrake}, \& {MacFadyen}}]{DeLaurentiis2024}
{DeLaurentiis}, S., {Haiman}, Z., {Westernacher-Schneider}, J.~R., {et~al.} 2024, arXiv e-prints, arXiv:2405.07897, \dodoi{10.48550/arXiv.2405.07897}

\bibitem[{{DiMatteo} {et~al.}(2023){DiMatteo}, {Ni}, {Chen}, {Croft}, \& {Pacucci}}]{DiMatteo2023}
{DiMatteo}, T., {Ni}, Y., {Chen}, N., {Croft}, R., \& {Pacucci}, F. 2023, in AAS/High Energy Astrophysics Division, Vol.~20, AAS/High Energy Astrophysics Division, 103A.04

\bibitem[{{Dittmann} \& {Ryan}(2021)}]{DittmannRyan2021}
{Dittmann}, A.~J., \& {Ryan}, G. 2021, \apj, 921, 71, \dodoi{10.3847/1538-4357/ac1bbd}

\bibitem[{Dittmann {et~al.}(2023)Dittmann, Ryan, \& Miller}]{Dittmann_2023}
Dittmann, A.~J., Ryan, G., \& Miller, M.~C. 2023, The Astrophysical Journal Letters, 949, L30, \dodoi{10.3847/2041-8213/acd183}

\bibitem[{{D'Orazio} \& {Charisi}(2023)}]{2023arXiv231016896D}
{D'Orazio}, D.~J., \& {Charisi}, M. 2023, arXiv e-prints, arXiv:2310.16896, \dodoi{10.48550/arXiv.2310.16896}

\bibitem[{{D'Orazio} \& {Duffell}(2021)}]{2021ApJ...914L..21D}
{D'Orazio}, D.~J., \& {Duffell}, P.~C. 2021, \apjl, 914, L21, \dodoi{10.3847/2041-8213/ac0621}

\bibitem[{{D'Orazio} {et~al.}(2024){D'Orazio}, {Duffell}, \& {Tiede}}]{DOrazio2024}
{D'Orazio}, D.~J., {Duffell}, P.~C., \& {Tiede}, C. 2024, \apj, 977, 244, \dodoi{10.3847/1538-4357/ad938b}

\bibitem[{{D'Orazio} \& {Haiman}(2017)}]{DOrazio2017}
{D'Orazio}, D.~J., \& {Haiman}, Z. 2017, \mnras, 470, 1198, \dodoi{10.1093/mnras/stx1269}

\bibitem[{D'Orazio {et~al.}(2013)D'Orazio, Haiman, \& MacFadyen}]{10.1093/mnras/stt1787}
D'Orazio, D.~J., Haiman, Z., \& MacFadyen, A. 2013, Monthly Notices of the Royal Astronomical Society, 436, 2997, \dodoi{10.1093/mnras/stt1787}

\bibitem[{{Duffell} {et~al.}(2024){Duffell}, {Dittmann}, {D'Orazio}, {Franchini}, {Kratter}, {Penzlin}, {Ragusa}, {Siwek}, {Tiede}, {Wang}, {Zrake}, {Dempsey}, {Haiman}, {Lupi}, {Pirog}, \& {Ryan}}]{2024ApJ...970..156D}
{Duffell}, P.~C., {Dittmann}, A.~J., {D'Orazio}, D.~J., {et~al.} 2024, \apj, 970, 156, \dodoi{10.3847/1538-4357/ad5a7e}

\bibitem[{D’Orazio {et~al.}(2016)D’Orazio, Haiman, Duffell, MacFadyen, \& Farris}]{D_Orazio_2016}
D’Orazio, D.~J., Haiman, Z., Duffell, P., MacFadyen, A., \& Farris, B. 2016, Monthly Notices of the Royal Astronomical Society, 459, 2379–2393, \dodoi{10.1093/mnras/stw792}

\bibitem[{Einstein(1916)}]{https://doi.org/10.1002/andp.19163540702}
Einstein, A. 1916, Annalen der Physik, 354, 769, \dodoi{https://doi.org/10.1002/andp.19163540702}

\bibitem[{Farris {et~al.}(2014)Farris, Duffell, MacFadyen, \& Haiman}]{10.1093/mnrasl/slu184}
Farris, B.~D., Duffell, P., MacFadyen, A.~I., \& Haiman, Z. 2014, Monthly Notices of the Royal Astronomical Society: Letters, 447, L80, \dodoi{10.1093/mnrasl/slu184}

\bibitem[{{Farris} {et~al.}(2015){Farris}, {Duffell}, {MacFadyen}, \& {Haiman}}]{FarrisDuffel2015}
{Farris}, B.~D., {Duffell}, P., {MacFadyen}, A.~I., \& {Haiman}, Z. 2015, \mnras, 447, L80, \dodoi{10.1093/mnrasl/slu184}

\bibitem[{{Franchini} {et~al.}(2024){Franchini}, {Bonetti}, {Lupi}, \& {Sesana}}]{Franchini2024}
{Franchini}, A., {Bonetti}, M., {Lupi}, A., \& {Sesana}, A. 2024, \aap, 686, A288, \dodoi{10.1051/0004-6361/202449206}

\bibitem[{{Garg} {et~al.}(2024{\natexlab{a}}){Garg}, {Tiede}, \& {D'Orazio}}]{Mudit2024}
{Garg}, M., {Tiede}, C., \& {D'Orazio}, D.~J. 2024{\natexlab{a}}, \mnras, 534, 3705, \dodoi{10.1093/mnras/stae2357}

\bibitem[{{Garg} {et~al.}(2024{\natexlab{b}}){Garg}, {Tiwari}, {Derdzinski}, {Baker}, {Marsat}, \& {Mayer}}]{2024MNRAS.528.4176G}
{Garg}, M., {Tiwari}, S., {Derdzinski}, A., {et~al.} 2024{\natexlab{b}}, \mnras, 528, 4176, \dodoi{10.1093/mnras/stad3477}

\bibitem[{Gommers {et~al.}(2024)Gommers, Virtanen, Haberland, Burovski, Weckesser, Reddy, Oliphant, Cournapeau, Nelson, alexbrc, Roy, Peterson, Polat, Wilson, endolith, Mayorov, van~der Walt, Brett, Laxalde, Larson, Sakai, Millman, Colley, Lars, peterbell10, Carey, van Mulbregt, Bowhay, eric jones, \& Striega}]{scipy_13352243}
Gommers, R., Virtanen, P., Haberland, M., {et~al.} 2024, scipy/scipy: SciPy 1.14.1, v1.14.1,  Zenodo, \dodoi{10.5281/zenodo.13352243}

\bibitem[{{Gong} {et~al.}(2021){Gong}, {Luo}, \& {Wang}}]{2021NatAs...5..881G}
{Gong}, Y., {Luo}, J., \& {Wang}, B. 2021, Nature Astronomy, 5, 881, \dodoi{10.1038/s41550-021-01480-3}

\bibitem[{{Gould} \& {Rix}(2000)}]{GouldRix2000}
{Gould}, A., \& {Rix}, H.-W. 2000, \apjl, 532, L29, \dodoi{10.1086/312562}

\bibitem[{{G{\"u}ltekin} \& {Miller}(2012)}]{2012ApJ...761...90G}
{G{\"u}ltekin}, K., \& {Miller}, J.~M. 2012, \apj, 761, 90, \dodoi{10.1088/0004-637X/761/2/90}

\bibitem[{{Guti{\'e}rrez} {et~al.}(2022){Guti{\'e}rrez}, {Combi}, {Noble}, {Campanelli}, {Krolik}, {L{\'o}pez Armengol}, \& {Garc{\'\i}a}}]{2022ApJ...928..137G}
{Guti{\'e}rrez}, E.~M., {Combi}, L., {Noble}, S.~C., {et~al.} 2022, \apj, 928, 137, \dodoi{10.3847/1538-4357/ac56de}

\bibitem[{Harris {et~al.}(2020)Harris, Millman, van~der Walt, Gommers, Virtanen, Cournapeau, Wieser, Taylor, Berg, Smith, Kern, Picus, Hoyer, van Kerkwijk, Brett, Haldane, del R{\'{i}}o, Wiebe, Peterson, G{\'{e}}rard-Marchant, Sheppard, Reddy, Weckesser, Abbasi, Gohlke, \& Oliphant}]{numpy}
Harris, C.~R., Millman, K.~J., van~der Walt, S.~J., {et~al.} 2020, Nature, 585, 357, \dodoi{10.1038/s41586-020-2649-2}

\bibitem[{Hirsh {et~al.}(2020)Hirsh, Price, Gonzalez, Ubeira-Gabellini, \& Ragusa}]{Hirsh2020}
Hirsh, K., Price, D., Gonzalez, J.-F., Ubeira-Gabellini, M., \& Ragusa, E. 2020, Monthly Notices of the Royal Astronomical Society, 498, 2936, \dodoi{10.1093/mnras/staa2536}

\bibitem[{Holz \& Hughes(2005)}]{Holz_2005}
Holz, D.~E., \& Hughes, S.~A. 2005, The Astrophysical Journal, 629, 15–22, \dodoi{10.1086/431341}

\bibitem[{{Kelley} {et~al.}(2017){Kelley}, {Blecha}, \& {Hernquist}}]{Kelley2017}
{Kelley}, L.~Z., {Blecha}, L., \& {Hernquist}, L. 2017, \mnras, 464, 3131, \dodoi{10.1093/mnras/stw2452}

\bibitem[{Khan {et~al.}(2012)Khan, Berentzen, Berczik, Just, Mayer, Nitadori, \& Callegari}]{Khan_2012}
Khan, F.~M., Berentzen, I., Berczik, P., {et~al.} 2012, The Astrophysical Journal, 756, 30, \dodoi{10.1088/0004-637X/756/1/30}

\bibitem[{King \& Pringle(2006)}]{10.1111/j.1745-3933.2006.00249.x}
King, A.~R., \& Pringle, J.~E. 2006, Monthly Notices of the Royal Astronomical Society: Letters, 373, L90, \dodoi{10.1111/j.1745-3933.2006.00249.x}

\bibitem[{{Kocsis} {et~al.}(2012){Kocsis}, {Haiman}, \& {Loeb}}]{2012MNRAS.427.2680K}
{Kocsis}, B., {Haiman}, Z., \& {Loeb}, A. 2012, \mnras, 427, 2680, \dodoi{10.1111/j.1365-2966.2012.22118.x}

\bibitem[{Krauth {et~al.}(2023)Krauth, Davelaar, Haiman, Westernacher-Schneider, Zrake, \& MacFadyen}]{10.1093/mnras/stad3095}
Krauth, L.~M., Davelaar, J., Haiman, Z., {et~al.} 2023, Monthly Notices of the Royal Astronomical Society, 526, 5441, \dodoi{10.1093/mnras/stad3095}

\bibitem[{{MacFadyen} \& {Milosavljevi{\'c}}(2008)}]{2008ApJ...672...83M}
{MacFadyen}, A.~I., \& {Milosavljevi{\'c}}, M. 2008, \apj, 672, 83, \dodoi{10.1086/523869}

\bibitem[{Mangiagli {et~al.}(2022)Mangiagli, Caprini, Volonteri, Marsat, Vergani, Tamanini, \& Inchauspé}]{Mangiagli_2022}
Mangiagli, A., Caprini, C., Volonteri, M., {et~al.} 2022, Physical Review D, 106, \dodoi{10.1103/physrevd.106.103017}

\bibitem[{Mastrobuono-Battisti {et~al.}(2024)Mastrobuono-Battisti, Seoane, i~Alfonso, Omarov, Yurin, Makukov, Omarova, \& Ogiya}]{MastrobuonoBattisti2024}
Mastrobuono-Battisti, A., Seoane, P.~A., i~Alfonso, M. J.~F., {et~al.} 2024, Prograde and retrograde stars in nuclear cluster mergers. Evolution of the supermassive black hole binary and the host galactic nucleus.
\newblock \doarXiv{2411.05063}

\bibitem[{Mayer(2013)}]{Mayer_2013}
Mayer, L. 2013, Classical and Quantum Gravity, 30, 244008, \dodoi{10.1088/0264-9381/30/24/244008}

\bibitem[{{Miller} \& {Krolik}(2013)}]{2013ApJ...774...43M}
{Miller}, M.~C., \& {Krolik}, J.~H. 2013, \apj, 774, 43, \dodoi{10.1088/0004-637X/774/1/43}

\bibitem[{Moody {et~al.}(2019)Moody, Shi, \& Stone}]{Moody_2019}
Moody, M. S.~L., Shi, J.-M., \& Stone, J.~M. 2019, The Astrophysical Journal, 875, 66, \dodoi{10.3847/1538-4357/ab09ee}

\bibitem[{Morais \& Giuppone(2012)}]{10.1111/j.1365-2966.2012.21151.x}
Morais, M. H.~M., \& Giuppone, C.~A. 2012, Monthly Notices of the Royal Astronomical Society, 424, 52, \dodoi{10.1111/j.1365-2966.2012.21151.x}

\bibitem[{Most \& Wang(2024)}]{most2024decouplingsupermassiveblackhole}
Most, E.~R., \& Wang, H.-Y. 2024, Decoupling of a supermassive black hole binary from its magnetically arrested circumbinary accretion disk.
\newblock \doarXiv{2410.23264}

\bibitem[{Muñoz {et~al.}(2019)Muñoz, Miranda, \& Lai}]{Muñoz_2019}
Muñoz, D.~J., Miranda, R., \& Lai, D. 2019, The Astrophysical Journal, 871, 84, \dodoi{10.3847/1538-4357/aaf867}

\bibitem[{Nixon(2012)}]{10.1111/j.1365-2966.2012.21072.x}
Nixon, C.~J. 2012, Monthly Notices of the Royal Astronomical Society, 423, 2597, \dodoi{10.1111/j.1365-2966.2012.21072.x}

\bibitem[{{Nixon} {et~al.}(2011){Nixon}, {Cossins}, {King}, \& {Pringle}}]{Nixon2011}
{Nixon}, C.~J., {Cossins}, P.~J., {King}, A.~R., \& {Pringle}, J.~E. 2011, \mnras, 412, 1591, \dodoi{10.1111/j.1365-2966.2010.17952.x}

\bibitem[{{Noble} {et~al.}(2012){Noble}, {Mundim}, {Nakano}, {Krolik}, {Campanelli}, {Zlochower}, \& {Yunes}}]{2012ApJ...755...51N}
{Noble}, S.~C., {Mundim}, B.~C., {Nakano}, H., {et~al.} 2012, \apj, 755, 51, \dodoi{10.1088/0004-637X/755/1/51}

\bibitem[{Paschalidis {et~al.}(2021)Paschalidis, Bright, Ruiz, \& Gold}]{Paschalidis_2021}
Paschalidis, V., Bright, J., Ruiz, M., \& Gold, R. 2021, The Astrophysical Journal Letters, 910, L26, \dodoi{10.3847/2041-8213/abee21}

\bibitem[{{Peters}(1964)}]{1964PhRv..136.1224P}
{Peters}, P.~C. 1964, Physical Review, 136, 1224, \dodoi{10.1103/PhysRev.136.B1224}

\bibitem[{Roedig \& Sesana(2014)}]{10.1093/mnras/stu194}
Roedig, C., \& Sesana, A. 2014, Monthly Notices of the Royal Astronomical Society, 439, 3476, \dodoi{10.1093/mnras/stu194}

\bibitem[{{Ruan} {et~al.}(2020){Ruan}, {Guo}, {Cai}, \& {Zhang}}]{2020IJMPA..3550075R}
{Ruan}, W.-H., {Guo}, Z.-K., {Cai}, R.-G., \& {Zhang}, Y.-Z. 2020, International Journal of Modern Physics A, 35, 2050075, \dodoi{10.1142/S0217751X2050075X}

\bibitem[{Schnittman(2010)}]{Schnittman_2010}
Schnittman, J.~D. 2010, The Astrophysical Journal, 724, 39, \dodoi{10.1088/0004-637X/724/1/39}

\bibitem[{Schnittman \& Krolik(2008)}]{Schnittman_2008}
Schnittman, J.~D., \& Krolik, J.~H. 2008, The Astrophysical Journal, 684, 835–844, \dodoi{10.1086/590363}

\bibitem[{{Shakura} \& {Sunyaev}(1973)}]{1973A&A....24..337S}
{Shakura}, N.~I., \& {Sunyaev}, R.~A. 1973, \aap, 24, 337

\bibitem[{{Siwek} {et~al.}(2023){Siwek}, {Weinberger}, \& {Hernquist}}]{Siwek2023}
{Siwek}, M., {Weinberger}, R., \& {Hernquist}, L. 2023, \mnras, 522, 2707, \dodoi{10.1093/mnras/stad1131}

\bibitem[{{Speri} {et~al.}(2024){Speri}, {Barsanti}, {Maselli}, {Sotiriou}, {Warburton}, {van de Meent}, {Chua}, {Burke}, \& {Gair}}]{EMRI_Tests_of_GR_Speri}
{Speri}, L., {Barsanti}, S., {Maselli}, A., {et~al.} 2024, arXiv e-prints, arXiv:2406.07607, \dodoi{10.48550/arXiv.2406.07607}

\bibitem[{{Springel} {et~al.}(2001){Springel}, {Yoshida}, \& {White}}]{GadgetSimulations}
{Springel}, V., {Yoshida}, N., \& {White}, S. D.~M. 2001, \na, 6, 79, \dodoi{10.1016/S1384-1076(01)00042-2}

\bibitem[{Tamanini {et~al.}(2016)Tamanini, Caprini, Barausse, Sesana, Klein, \& Petiteau}]{Tamanini_2016}
Tamanini, N., Caprini, C., Barausse, E., {et~al.} 2016, Journal of Cosmology and Astroparticle Physics, 2016, 002, \dodoi{10.1088/1475-7516/2016/04/002}

\bibitem[{{Tanaka} {et~al.}(2010){Tanaka}, {Haiman}, \& {Menou}}]{Tanaka_BirthofaQuasar}
{Tanaka}, T., {Haiman}, Z., \& {Menou}, K. 2010, \aj, 140, 642, \dodoi{10.1088/0004-6256/140/2/642}

\bibitem[{Tanaka \& Haiman(2013)}]{Tanaka2013ElectromagneticSO}
Tanaka, T.~L., \& Haiman, Z. 2013, Classical and Quantum Gravity, 30.
\newblock \url{https://api.semanticscholar.org/CorpusID:118845333}

\bibitem[{Tang {et~al.}(2018)Tang, Haiman, \& MacFadyen}]{Tang_2018}
Tang, Y., Haiman, Z., \& MacFadyen, A. 2018, Monthly Notices of the Royal Astronomical Society, 476, 2249–2257, \dodoi{10.1093/mnras/sty423}

\bibitem[{Tiede \& D’Orazio(2023)}]{Tiede_2023}
Tiede, C., \& D’Orazio, D.~J. 2023, Monthly Notices of the Royal Astronomical Society, 527, 6021–6037, \dodoi{10.1093/mnras/stad3551}

\bibitem[{{Tiede} {et~al.}(2020){Tiede}, {Zrake}, {MacFadyen}, \& {Haiman}}]{2020ApJ...900...43T}
{Tiede}, C., {Zrake}, J., {MacFadyen}, A., \& {Haiman}, Z. 2020, \apj, 900, 43, \dodoi{10.3847/1538-4357/aba432}

\bibitem[{{Tiede} {et~al.}(2024){Tiede}, {Zrake}, {MacFadyen}, \& {Haiman}}]{2024arXiv241003830T}
---. 2024, arXiv e-prints, arXiv:2410.03830.
\newblock \doarXiv{2410.03830}

\bibitem[{Van~Rossum \& Drake(2009)}]{python}
Van~Rossum, G., \& Drake, F.~L. 2009, Python 3 Reference Manual (Scotts Valley, CA: CreateSpace)

\bibitem[{Virtanen {et~al.}(2020)Virtanen, Gommers, Oliphant, Haberland, Reddy, Cournapeau, Burovski, Peterson, Weckesser, Bright, {van der Walt}, Brett, Wilson, Millman, Mayorov, Nelson, Jones, Kern, Larson, Carey, Polat, Feng, Moore, {VanderPlas}, Laxalde, Perktold, Cimrman, Henriksen, Quintero, Harris, Archibald, Ribeiro, Pedregosa, {van Mulbregt}, \& {SciPy 1.0 Contributors}}]{2020SciPy-NMeth}
Virtanen, P., Gommers, R., Oliphant, T.~E., {et~al.} 2020, Nature Methods, 17, 261, \dodoi{10.1038/s41592-019-0686-2}

\bibitem[{Wagg {et~al.}(2024)Wagg, Broekgaarden, \& Gültekin}]{software-citation-station-zenodo}
Wagg, T., Broekgaarden, F., \& Gültekin, K. 2024, TomWagg/software-citation-station: v1.2, v1.2,  Zenodo, \dodoi{10.5281/zenodo.13225824}

\bibitem[{{Wagg} \& {Broekgaarden}(2024)}]{software-citation-station-paper}
{Wagg}, T., \& {Broekgaarden}, F.~S. 2024, arXiv e-prints, arXiv:2406.04405.
\newblock \doarXiv{2406.04405}

\bibitem[{Wang {et~al.}(2022)Wang, Songsheng, Li, \& Du}]{Wang_2022}
Wang, J.-M., Songsheng, Y.-Y., Li, Y.-R., \& Du, P. 2022, Monthly Notices of the Royal Astronomical Society, 518, 3397–3406, \dodoi{10.1093/mnras/stac3266}

\bibitem[{Westernacher-Schneider {et~al.}(2022)Westernacher-Schneider, Zrake, MacFadyen, \& Haiman}]{Westernacher_Schneider_2022}
Westernacher-Schneider, J.~R., Zrake, J., MacFadyen, A., \& Haiman, Z. 2022, Physical Review D, 106, \dodoi{10.1103/physrevd.106.103010}

\bibitem[{{Williamson} {et~al.}(2022){Williamson}, {B{\"o}sch}, \& {H{\"o}nig}}]{Williamson2022}
{Williamson}, D.~J., {B{\"o}sch}, L.~H., \& {H{\"o}nig}, S.~F. 2022, \mnras, 510, 5963, \dodoi{10.1093/mnras/stab3792}

\bibitem[{Zrake {et~al.}(2024)Zrake, Clyburn, \& Feyan}]{zrake2024changinglookinspiralstrendsswitches}
Zrake, J., Clyburn, M., \& Feyan, S. 2024, Changing-Look Inspirals: Trends and Switches in AGN Disk Emission as Signposts for Merging Black Hole Binaries.
\newblock \doarXiv{2410.04961}

\bibitem[{{Zrake} \& {MacFadyen}(2024)}]{2024ascl.soft08004Z}
{Zrake}, J., \& {MacFadyen}, A. 2024, {Sailfish: GPU-accelerated grid-based astrophysics gas dynamics code}, Astrophysics Source Code Library, record ascl:2408.004

\bibitem[{{Zrake} {et~al.}(2021){Zrake}, {Tiede}, {MacFadyen}, \& {Haiman}}]{2021ApJ...909L..13Z}
{Zrake}, J., {Tiede}, C., {MacFadyen}, A., \& {Haiman}, Z. 2021, \apjl, 909, L13, \dodoi{10.3847/2041-8213/abdd1c}

\end{thebibliography}

\appendix
\section{Cavity Measurements}
\label{Appendix}
Following \cite{Dittmann_2023}, we define the central cavity to be the inner region of the disk with surface density less than $\Sigma_\mathrm{cav} = 0.2 \Sigma_0$. To quantify the size of this region, we begin by constructing an isocontour of the surface density with value $\Sigma_\mathrm{cav}$. Along this contour, we sample $500$ points (of the form $\mathbf{x}_i = x_i\hat{x} + y_i\hat{y}$ for $1\leq i \leq 500$) which we seek to fit with an ellipse of semi-major axis $a_\mathrm{c}$, eccentricity $e_\mathrm{c}$ and argument of pericenter $\omega_\mathrm{c}$. Next, by defining the directed distance between points $i$ and $j$ as $\mathbf{d}_\mathrm{ij}  \equiv \mathbf{x}_i - \mathbf{x}_j$, the cavity semi-major axis can thus be identified as,
\begin{equation}
    a_\mathrm{c} = \frac{1}{2}\mathrm{max}\left(|\mathbf{d}_\mathrm{ij}|\right),
\end{equation}
corresponding to the maximum distance between any two points. We denote these two maximally distanced points $\mathbf{x}_-$ and $\mathbf{x}_+$, for which the argument of pericenter can be found as the relative angle between the two,
\begin{equation}
    \omega_\mathrm{c} = \arctan\left[\frac{(\mathbf{x}_+ - \mathbf{x}_-) \cdot \hat{y}}{(\mathbf{x}_+ - \mathbf{x}_-) \cdot \hat{x}}\right].
\end{equation}
To calculate the semi-minor axis of the ellipse, we require the maximum distance of two points in a direction orthogonal to $\omega_c$. We thus define the unit vector orthogonal to the argument of periapsis $\hat{n}$ as
\begin{equation}
    \hat{n} = -\sin(\omega_\mathrm{cav})\hat{x} + \cos(\omega_\mathrm{cav})\hat{y},
\end{equation}
for which, the projected distance along this direction is given as $n_{ij} = \hat{n}\cdot \mathbf{d}_{ij}$. Hence, the semi-minor axis of the ellipse $b_\mathrm{c}$ is given as
\begin{equation}
    b_\mathrm{cav} = \frac{1}{2}\mathrm{max}{(|n_{ij}|)}.
\end{equation}
Finally, using the semi-major and semi-minor axes of the cavity, we calculate its eccentricity as
\begin{equation}
    e_\mathrm{c} = \sqrt{1-\frac{b_\mathrm{c}^2}{a_\mathrm{c}^2}}.
\end{equation}

\end{document}